\documentclass[11pt]{article}
\pdfoutput=1
\usepackage{amsmath,amssymb,amsfonts,amsxtra, mathrsfs, makeidx,graphics,graphicx,amsthm,epsfig,jheppub1}

\newcommand{\beq} {\begin{equation}}
\newcommand{\eeq} {\end{equation}}
\newcommand{\beqa} {\begin{eqnarray}}
\newcommand{\eeqa} {\end{eqnarray}}

\newcommand{\morder}[1]{{\cal O}\left(#1 \right)}

\newcommand{\im}{\mathrm{Im}\,}
\newcommand{\re}{\mathrm{Re}\,}

\newcommand{\be}{\begin{equation}}
\newcommand{\ee}{\end{equation}}
\newcommand{\bea}{\begin{eqnarray}}
\newcommand{\eea}{\end{eqnarray}}

\newcommand{\ba}{\begin{array}}
\newcommand{\ea}{\end{array}}
\newcommand{\bi}{\begin{itemize}}
\newcommand{\ei}{\end{itemize}}
 \def\bea#1\eea{\allowdisplaybreaks \begin{align}#1\end{align}}
\newcommand{\ben}{\begin{enumerate}}
\newcommand{\een}{\end{enumerate}}
\newcommand{\bean}{\begin{eqnarray*}}
\newcommand{\eean}{\end{eqnarray*}}
\newcommand{\eref}[1]{(\ref{#1})}

\newcommand{\nn}{\nonumber}

\newcommand{\comment}[1]{}

\newcommand{\CD}{{\cal D}}

\newcommand{\CN}{{\cal N}}

\newcommand{\CR}{{\cal R}}
\newcommand{\CP}{{\cal P}}
\newcommand{\CH}{{\cal H}}

\newcommand{\CI}{{\cal I}}

\newcommand{\ie}{{\it i.e.}}
\newcommand{\eg}{{\it e.g.}}

\newcommand{\ud}{\mathrm{d}}

\newcommand{\sv}{\textsf{v}}

\newcommand{\tv}{\textsf{v}}

\newcommand{\sqx}{\sqrt{(b-x)(x-a)}}
\newcommand{\sqy}{\sqrt{(b-y)(y-a)}}

\newcommand{\sig}{\sigma}

\title{Moduli space of supersymmetric QCD\newline
in the Veneziano limit}

\author[a]{Yang Chen,}
\author[b,c]{Niko Jokela,}
\author[d]{Matti J\"arvinen,}
\author[e]{and Noppadol Mekareeya} 

\affiliation[a]{Faculty of Science and Technology, Department of Mathematics, \\
University of Macau, Av. Padre Tom\'as Pereira, Taipa Macau, China}
\affiliation[b]{Departamento de  F\'\i sica de Part\'\i culas, \\
Universidade de Santiago de Compostela} 
\affiliation[c]{Instituto Galego de F\'\i sica de Altas Enerx\'\i as (IGFAE), \\
E-15782, Santiago de Compostela, Spain} 
\affiliation[d]{Crete Center for Theoretical Physics, \\
Department of Physics, University of Crete, 71003 Heraklion, Greece} 
\affiliation[e]{Max-Planck-Institut f\"ur Physik (Werner-Heisenberg-Institut), \\
F\"ohringer Ring 6, 80805 M\"unchen, Deutschland}
\emailAdd{ yangbrookchen@yahoo.co.uk}
\emailAdd{ niko.jokela@usc.es}
\emailAdd{ mjarvine@physics.uoc.gr}
\emailAdd{noppadol@mpp.mpg.de}
\abstract{We study the moduli space of $4d$ $\CN=1$ supersymmetric QCD in the Veneziano limit using Hilbert series.  In this limit, the numbers of colours and flavours are taken to be large with their ratio fixed.  It is shown that the Hilbert series, which is a partition function of an ensemble of gauge invariant quantities parametrising the moduli space, can also be realised as a partition function of a system of interacting Coulomb gas in two dimensions. In the electrostatic equilibrium, exact and asymptotic analyses reveal that such a system exhibits two possible phases.  Physical quantities, such as charge densities, free energies, and Hilbert series, associated with each phase, are computed explicitly and discussed in detail.  We then demonstrate the existence of the third order phase transition in this system.}

\begin{document}
\setcounter{tocdepth}{2}
\maketitle

\section{Introduction and summary}
Hilbert series of a supersymmetric gauge theory is a partition function of the system of gauge invariant quantities parametrising the moduli space of vacua.  It can be viewed as a trace over the space of gauge invariant objects parametrising the vacua of the theory, counting with respect to a certain $U(1)$ global symmetry \cite{Pouliot:1998yv,Romelsberger:2005eg,Hanany:2006uc,Benvenuti:2006qr,Feng:2007ur,Gray:2008yu,Hanany:2008kn, Hanany:2008qc, Davey:2009sr}.  In the context of 4d $\CN=1$ supersymmetric QCD (SQCD), these gauge invariant objects are combinations of quark and antiquark superfields, known as mesons and baryons \cite{Gray:2008yu, Hanany:2008kn}.  It is convenient to take the $U(1)$ global symmetry to be proportional to the $R$-symmetry such that the quarks and antiquarks carry a unit charge.

In computing the partition function, we need to introduce a bookkeeping variable of the $U(1)$ global charge; this is known as the fugacity denoted by $t$, such that $0<t<1$.  The Hilbert series can be computed in two steps: 
\ben
\item Compute the generating function of symmetric functions of the quark and antiquark superfields.  This is a rational function of $t$ and Cartan variables of the gauge group $G$, denoted by $z_a$ for $a=1,2,\ldots, {{\rm rank}~G}$. 
\item Integrate this rational function over the Haar measure of the gauge group $G$.  This step restricts the generating function in Step 1 to count the gauge invariant combinations of quarks and antiquarks, namely mesons and baryons.   
\een
Since the Haar measure involves integrals along the contours $|z_a|=1$ for $a=1,2, \ldots,  {\rm rank}~G$, the Hilbert series can be written in terms of nested multi-contour integrals over a certain rational function.  These expressions are explicitly written in (4.9) of \cite{Gray:2008yu}, (3.14) of \cite{Hanany:2008kn}, (3.6) of \cite{Chen:2011wn}, (2.8), (3.3), (4.4) of \cite{Basor:2011da}, and (5), (8) of \cite{Jokela:2011vg}.

In \cite{Chen:2011wn}, \cite{Basor:2011da}, and \cite{Jokela:2011vg}, it was shown that these multi-contour integrals can be recast in terms of determinants of Toeplitz matrices for unitary and special unitary gauge groups, and determinants of Hankel matrices for the special orthogonal and symplectic gauge groups.  Writing these integrals in terms of determinants has several advantages: it allows for exact expressions for the Hilbert series for arbitrary number of colours and flavours, as well as various asymptotic results when the number of colours and flavours are large.  

One of the asymptotic limits considered in this paper is when the number of colours $N_c$ and the number of flavours $N_f$ tend to infinity, with their ratio kept fixed and finite.  This is known as the {\it Veneziano limit} \cite{Veneziano:1976wm}.   Historically, this limit has received a lot of interest in QCD due to the interesting topological nature of Feynman diagrams.  The main goal of this paper is to study the vacuum moduli space of SQCD via analyzing  
the Hilbert series in this asymptotic limit.

Another motivation to study the Veneziano limit of SQCD is that given the ratio $N_f/N_c$ in a certain range, the theory may exhibit a number of physically interesting aspects.  Let us draw some examples from  \cite{Seiberg:1994bz,Intriligator:1995id,Intriligator:1995ne} as follows:
\bi
\item For $SU(N_c)$, $SO(N_c)$, and $Sp(N_c)$ gauge groups, with respectively $N_f \geq N_c+2$, $N_f \geq N_c+3$, and $N_f \geq N_c+3$,  each of these theories not only possesses a degenerate vacua at the quantum level,\footnote{It was shown in \cite{Seiberg:1994bz,Intriligator:1995id,Intriligator:1995ne} that, for $SU(N_c)$ SQCD with  $N_f \leq N_c-1$, $SO(N_c)$ SQCD with $N_f \leq N_c-5$, and $Sp(N_c)$ SQCD with $N_f \leq N_c$, the moduli spaces are totally lifted by dynamical generated superpotential and hence there are no supersymmetric vacua.} but its classical moduli space also receives no quantum corrections.  Hence, the Hilbert series computed for the classical moduli space as described above still provides a {\it valid} description of the quantum moduli space.  
\item In these ranges of $N_f$ and $N_c$, the theory possesses a dual description.  This phenomenon is known as the {\it Seiberg duality} \cite{Seiberg:1994pq}. The Hilbert series will therefore also describe the quantum moduli space of the dual theory.
\item If $N_f/N_c$ is fixed in a certain interval, known as the {\it conformal window}, the theory has a non-trivial infrared fixed point.  For $SU(N_c)$, $Sp(N_c)$, and $SO(N_c)$ gauge groups,  the conformal windows are $\frac{3}{2}N_c < N_f < 3N_c$, $\frac{3}{2}(N_c+1) < N_f < 3(N_c+1)$, and $\frac{3}{2}(N_c-2) < N_f < 3(N_c-2)$, respectively.  Within this interval, the Hilbert series describes the vacua of the theory at the conformal fixed point. 
\ei

The moduli space of $U(N_c)$ and $SU(N_c)$ SQCD with $N_f$ flavours in this asymptotic limit was studied in detail in \cite{Chen:2011wn,Jokela:2011vg}.  In this paper, we focus on SQCD with $SO(N_c)$ and $Sp(N_c)$ gauge groups.

Let us now present an overview of the approach we use in this paper.  Regarding the Hilbert series as a partition function of the system of gauge invariant objects, we can read off the Hamiltonian of this system from the multi-integral form of the Hilbert series.  Such a Hamiltonian in the Veneziano limit can be interpreted as the Hamiltonian of a system of Coulomb gas or log-gas in two dimensions.  The charged particles are constrained to reside in a one-dimensional interval inside the two-dimensional space.  The charges in this interval interact among themselves and are also subject to the interactions due to external fields; the latter is determined by the number of ratio $N_f/N_c$ and the gauge group.  By solving the equilibrium condition for this system, we can compute the charge density within the aforementioned one-dimensional interval.  This quantity exhibits two phases of the system, dubbed the {\it gapless} and the {\it gapped} phases, with the order parameter being the size of the gap. 
Which of these two phases the system belongs depends on the ratio $N_f/N_c$ and on the value of the fugacity $t$ in the Hilbert series.    The free energy of the system can be computed in each phase, from which the Hilbert series can be deduced.  By examining the behavior of the free energy at the transition point between these two phases, we demonstrate the possibility of the third order phase transition, 
reminiscient of \cite{Gross:1980he,Wadia:2012fr}. This phase transition was uncovered for the $U(N_c)$ and $SU(N_c)$ gauge groups in \cite{Jokela:2011vg}.

\subsection*{Outline and key results}
This paper is divided  
into four main parts.  The first part, containing Section \ref{sec:CGpert}, develops the connection between Hilbert series and the partition function of the Coulomb gas in two dimensions, which was
established for the $U(N_c)$ and $SU(N_c)$ gauge groups in \cite{Jokela:2011vg}. The basic connection goes back to the seminal works of 
Wigner and Dyson \cite{Wigner,dyson:140,dyson:157,dyson:166,dyson:701,mehta:713}. While the electrostatic analogues have a long 
history in string theory \cite{Fairlie:1970tc,Gross:1987kza} (see also \cite{Barbon:1996ie,Bachas:1999tv}), they
have received considerable attention only relatively recently (see \eg, \cite{Balasubramanian:2004fz, Marino:2004eq,Jokela:2005ha,Balasubramanian:2006sg,Jokela:2007wi,Hutasoit:2007wj,Jokela:2007yc,Jokela:2007dq,Jokela:2008zh,Jokela:2009gc,Jokela:2009fd,Jokela:2010zp, Drukker:2010nc,Marino:2012zq,Jain:2013py}).  
By exploiting the Coulomb gas picture we coarse-grain the Hamiltonian of the Coulomb gas system in the Veneziano limit and determine the electrostatic equilibrium condition.   

We then proceed to the second part of the paper, consisting of Sections \ref{sec:exactv} and \ref{sec:pertsolutions}.  In Section \ref{sec:exactv}, we generalise the method of \cite{BasorChen} to find exact solutions to this equilibrium condition, from which expressions for the charge density and the free energy are derived.  These exact expressions involve certain variables determined by degree six polynomial equations. Thus, in order to make these exact expressions explicit in the Veneziano limit, we need to solve such equations perturbatively in $1/N_c$ (with the ratio $N_f/N_c$ kept fixed) and this is the main goal of Section \ref{sec:pertsolutions}.  These perturbative solutions suggest that the Coulomb gas system at the equilibrium exhibits two phases.  The charge density, the free energy, and the Hilbert series for each phase are computed.  The agreement between the Hilbert series computed in this paper and known results (or limits of known results) presented in \cite{Hanany:2008kn, Basor:2011da} 
justifies the validity of our perturbative solutions.

In the third part of the paper, namely Section \ref{sec:pertanalysis}, we present another approach to derive the key results for the Hilbert series. We apply directly a perturbative analysis to the Coulomb gas system in the Veneziano limit. We solve the electrostatic equilibrium condition at leading order -- the next to leading order correction to the free energy, and consequently to the Hilbert series, then immediately follow by the standard perturbation theory approach. This is enough to reproduce the main results of Section~\ref{sec:pertsolutions}.

In the last part of the paper, we examine the behaviour of the Coulomb gas system near the transition point between the two phases.  The order parameter and the free energy are investigated.  We demonstrate the situation in which the system exhibits the third order phase transition.

We summarise the key points of this paper below.
\bi
\item  The two phases of the system are discussed and summarised above \eref{criticalT} and below \eref{eq:poteq2}.  The equilibrium configurations of the Coulomb gas for each phase are solved explicitly in Sections~\ref{sec:gaplesssoln},~\ref{sec:gappedphase}, and~\ref{sec:pertanalysis}.
\item  The Hilbert series computed in the gapless phase are given by \eref{HSgapless} and those computed in the gapped phase (in a certain limit) are given by \eref{HSCngap}--\eref{HSDngap}.
\item  The existence of the third order phase transition is shown in \eref{3rdorder}.
\ei

\section{The Hilbert series of SQCD and the Coulomb gas} \label{sec:CGpert}
The Hilbert series for SQCD with classical gauge groups $SO(2n+1)$, $SO(2n)$, and $Sp(n)$ with $N_f$ flavours can be written as \cite{Basor:2011da}
\bea \label{HS}
 g_{N_f,G}(t) = \begin{cases} 
 \frac{2^{n^2}}{(1-t)^{N_f}(2 t)^{N_f n}} {\CD}_{G}(n,N_f) &\quad \text{for~$G=SO(2n+1)=B_n$} \\
 \frac{2^{(n-1)^2+n}}{(2 t)^{N_f n}} {\CD}_{G}(n,N_f) &\quad \text{for~$G=SO(2n)=D_n$} \\
 \frac{2^{n^2+n}}{(2 t)^{2 N_f n}} {\CD}_{G} (n,2N_f) &\quad \text{for~$G=Sp(n)=C_n$}\ , \\
\end{cases}
\eea
where ${\CD}_G(n,N_f)$ is defined by
\bea
{\CD}_G(n,N_f)  &= \frac{1}{(2\pi)^n n!} \int_{-1}^1 \ud x_1 \cdots \int_{-1}^1 \ud x_n \prod_{1 \leq a<b \leq n} (x_a-x_b)^2 \prod_c \frac{w_G(x_c)}{\left(-x_c+\frac{1+t^2}{2 t}\right)^{N_f}} \nn \\
&=  \frac{1}{(2\pi)^n n!}\int_{(-1,1)^n}\:\: \prod_{1\leq a<b\leq
n}(x_a-x_b)^{2}\prod_{c=1}^{n} w_G(x_c)   \;
e^{-N_f\log(T-x_c)}  \ud x_c~,
\eea
where 
\bea
T := \frac{1+t^2}{2t} \qquad  \text{with}~T>1~, \label{Tandt}
\eea
and $w_G(x)$ is known as the {\it Jacobi weight} defined by
\bea
w_G(x) = (1-x)^\alpha (1+x)^\beta~, \label{jacobiweight}
\eea
with
\bea \label{alphabeta}
(\alpha, \beta) = \begin{cases} (1/2,-1/2) &\quad \text{for $G=SO(2n+1)=B_n$} \\
(-1/2,-1/2) &\quad \text{for $G=SO(2n)=D_n$} \\
(1/2,1/2) &\quad \text{for $G=Sp(n)=C_n$}~. \\
\end{cases}
\eea
Note that, as discussed in \cite{Basor:2011da}, ${\CD}_G(n,N_f)$ can be written as a determinant of a Hankel matrix.

We regard the Hilbert series and hence ${\CD}_G(n,N_f)$ as a partition function of the system consisting of gauge invariant quantities parametrising the moduli space. Let us rewrite ${\CD}_G(n,N_f)$ as
\bea
{\CD}_G(n,N_f) &= \frac{1}{(2\pi)^n n!}\int_{-1}^1 \ud x_1 \cdots \int_{-1}^1 \ud x_n ~e^{-n^2 H(T,x)}~, \label{defDG}
\eea
and interpret $H$ as the {\it Hamiltonian} or the {\it energy} of the system given by
\bea
H(T,x) &=  -2 n^{-2} \sum_{1 \leq a<b \leq n} \log |x_a-x_b| - n^{-2} \sum_{c=1}^n \log w_G(x_c) + n^{-2} N_f \sum_{c=1}^n \log|T-x_c| \nn \\
 &= n^{-2} \left[ -\sum_{1 \leq a \ne b \leq n} \log |x_a-x_b| - \sum_{c=1}^n \log w_G(x_c) + N_f \sum_{c=1}^n \log|T-x_c| \right]~.
\eea

The energy of this system can be identified as the potential energy of the two-dimensional Coulomb gas, subject to certain external interactions.  Consider identical charged particles, each with charge $+1$, located at the position $x_1, \ldots, x_n$ within an interval $(-1,1)$ on the $x$-axis in a two-dimensional space. The physical interpretation of each term in the square bracket is as follows:
\bi
\item The first term is the Coulomb interaction in two spatial dimensions between the particles within the interval $(-1,1)$.
\item The third term denotes the total Coulomb interaction between the charge particle at $x=x_c$, with  $x_c \in (-1,1)$, and an external particle with the charge $-N_f$ located at the position $x =T>1$.
\item The second term indicates the total interaction between the charged interval with a certain external field determined by the Jacobi weight $w_G(x)$.  Using the explicit form \eref{jacobiweight}, this external field is sourced by the extra particles located at $x = \pm 1$ with the charges $\alpha$ and $\beta$.
\ei

\subsection{The continuum limit} 
Let us consider the continuum limit $n \rightarrow \infty$.  We approximate $H$ by using the connection to the continuum Coulomb gas (see the books \cite{mehta,forrester}) as follows:
\bea
H \sim \CH &:=  \frac{N_f}{n} \int_{a}^{b} \: \sig(x)\log(T-x)\ud x-  n^{-1} \int_{a}^{b}\sig(x)\log w_G(x) \ud x \nn \\
& \quad - \int_{a}^{b} \ud x \int_{a}^{b}  \: \ud x' \: \sig(x)\log|x-x'|\sig(x')~, \label{Hamil}
\eea
where $\sigma(x)$ is the density of the charged particles supported on the interval $a \leq x \leq b$, satisfying the normalisation condition
\bea
\int_{a}^{b}\sig(x)\ud x=1~, \label{normsig}
\eea
and the integration limits $a$ and $b$ are such that 
\bea
-1 \leq a < b \leq 1~.
\eea
The Hilbert series calculated from the continuum approximation (\ref{Hamil}) turns out to be correct at next-to-leading order; see 
subsections \ref{sec:largen} and \ref{sec:Hilb} for comments.
Because the (discrete) Coulomb gas contains only positive charges, $\sigma(x)$ is constrained to be non-negative throughout the interval $[a,b]$.

In the end we will be interested in the Veneziano limit, i.e., we will take $n, \: N_f \rightarrow \infty$ with their ratio kept fixed and of order $1$. In view of this, and in order to take into account the factor of two appearing for $Sp(n)$ in the integral $\CD_G$ of~\eqref{HS}, it is convenient to define\footnote{For $SU(n)$, it is natural to define $\CR = 2N_f/n$.  As can be seen subsequently, with this definition of $\CR$, quantities such as the critical temperature $t_c$ given by \eref{criticalT} agree with that of \cite{Jokela:2011vg}.}
\bea \label{defR}
\CR = \begin{cases} \frac{N_f}{n} \qquad &\text{for $G=B_n,D_n$}~, \\
\frac{2N_f}{n} \qquad &\text{for $G=C_n$}~. \end{cases}  
\eea
Note that in this limit, the conformal window corresponds to the interval
\bea
3 < \CR <6~, \label{confwindow}
\eea
independent of $G$.

We can thus write \eref{Hamil} for each gauge group as
\bea
\CH = \int_a^b \ud x ~\sig(x)  \left[ \textsf{v}(x)- \int_{a}^{b}\log|x-y|\sig(y)\ud y \right] ~, \label{freeen2}
\eea 
where $\textsf{v}(x)$ is the {\it external potential} given by
\bea
\textsf{v}(x) &= \CR \;\log(T-x)- n^{-1} \log w_G(x) \nn \\
&= \CR \;\log(T-x)- n^{-1} \left[\alpha \log(1-x) + \beta \log(1+x) \right]~. \label{extpot}
\eea

\subsubsection{The equilibrium condition} 
In order to determine the configuration with the minimum energy (\ie, the equilibrium condition), we combine the following variation
\bea
\delta \CH = \int_{a}^{b} \ud x \left[ \textsf{v}(x)- 2\int_{a}^{b}\log|x-y|\sig(y)\ud y \right] \delta \sigma(x)~.
\eea
with the variation of the constraint \eref{normsig} multiplied by a Lagrange multiplier $-A$,
\bea
-A \int_a^b \ud x~ \delta \sigma(x) =0~, \qquad A = \text{constant}
\eea
and equate to zero:
\bea
0 = \int_{a}^{b} \ud x \left[ \textsf{v}(x)-2\int_{a}^{b}\log|x-y|\sig(y)\ud y -A\right] \delta \sigma(x)~.
\eea
Therefore, we arrive at the equilibrium condition
\bea
\textsf{v}(x)-2\int_{a}^{b}\log|x-y|\sig(y)\ud y = \text{constant} = A ~. \label{constantA}
\eea
This relation indicates that the electrostatic potential, given as the sum of the internal Coulomb potential due to the interacting charges within the interval $(a,b)$ and the external potential, is constant at the equilibrium for any point $x \in (a,b)$.  

We shall henceforth assume that the system is always at the equilibrium.  To avoid cumbersome notation, let us slightly make abuse of notation by denoting by $\CH$ the energy of the system at the equilibrium.  From \eref{freeen2}, $\CH$ satisfies
\bea
2\CH = A+I~,
\eea
where
\bea
I := \int_a^b \ud x ~{\sf v}(x) \sigma(x)~. \label{freeenc}
\eea

\subsubsection{The Hilbert series at large $n$}\label{sec:largen}
The partition function ${\CD}_G(n,N_f)$ of~\eref{defDG}, and consequently the Hilbert series through~\eqref{HS}, can be approximated in terms of the equilibrium Hamiltonian (see, \eg, \cite{Szego,ChenLawrence,BasorChen}):
\bea
{\CD}_G(n,N_f) 
 \sim e^{-n^2 \CH}~.
\eea
We can thus interpret $n^2 \CH$ as the {\it free energy} of the system,
\bea \label{freeenA}
F := n^2 \CH = \frac{1}{2} n^2 (A+I)~,
\eea
which satisfies
\bea 
 F \sim  -\log {\CD}_G(n,N_f) ~.\label{FandD}
\eea

In order to derive the result~\eqref{FandD} we have approximated the Hamiltonian by using the continuum Coulomb gas in (\ref{Hamil}), and only 
included the equilibrium (saddle-point) contribution. It is important to control the size of the corrections due to these approximations.
Earlier analytic and numerical analyses of similar systems suggest that this result is accurate up to, and including, 
leading and next-to-leading\footnote{In order for the relation~\eqref{FandD} to hold at next-to-leading order, 
it is essential that we are working at a critical temperature of the Coulomb gas. At noncritical temperatures, 
there are finite (but easily calculable) corrections even at next-to-leading order (see, e.g., Appendix~A in~\cite{Jokela:2010cc}).} 
order terms in the 't Hooft (large $n$)~\cite{BasorChen} and Veneziano~\cite{Jokela:2011vg} limits. 
It is possible to check this numerically by comparing our results for the free energy in Section~\ref{sec:exactv} to the exact 
results in terms of Hankel determinants given in~\cite{Basor:2011da}.
We also provide a non-trivial consistency check of the next-to-leading order results in subsection \ref{sec:Hilb} in the limit $t\to 1$.

\section{Exact results in the continuum limit} \label{sec:exactv}
Having obtained the equilibrium condition \eref{constantA}, we proceed further in Section \ref{sec:chargedensityexact} by solving this condition for the charge density $\sigma$ in an exact way.  Subsequently, in Section \ref{sec:freeenergy}, we use this exact expression and \eref{freeenA} to obtain the expression for the free energy $F$.

As we shall see below, the exact expressions for $\sigma$ and $F$ are functions of the upper and lower limits $a$ and $b$.  The latter are determined by degree six equations.  In Section \ref{sec:pertsolutions}, we find perturbative solutions to such equations in the Veneziano limit.

\subsection{The charge density of the Coulomb gas} \label{sec:chargedensityexact}
The external force acting on the charged interval can be calculated by differentiating \eref{constantA} with respect to $x$.  This yields
\bea
\textsf{v}'(x) -2 \CP \int_{a}^{b}\frac{\sig(y)}{x-y}\ud y=0~, \label{extforce}
\eea 
where $\CP$ denotes the principal value.  We aim to compute the charge density $\sigma(x)$ as a solution to this integral equation, with the normalisation according to \eref{normsig}.\footnote{An alternative method of solving the equilibrium configuration is discussed in Appendix~\ref{app:confmap}.}  This can be achieved using the knowledge from the theory of singular integral equations.\footnote{See, \eg,~ Eqs. (25) and (26) in \S 26 of \cite{mikhlin1964integral} and Eq. (89.16) of \cite{muskhelishvili2008singular} for further details.}  

We follow closely the analysis of \cite{BasorChen}. First taking $\alpha,\beta>0$, we expect that for the unique physical, positive definite solution, the density is supported in the interval $(a,b)$ with $-1<a<b<1$ and satisfies the boundary conditions
\bea \label{assumpBC}
\sigma(x) \rightarrow 0 \quad \text{as} \quad x \rightarrow a^+ \quad \text{or} \quad x \rightarrow b^-~.
\eea
The solution subject to these boundary conditions is\footnote{The analysis in \cite{BasorChen} corresponds to our computation with $\CR=0$.  The contribution from a non-zero $\CR$ in our analysis comes in as the first term of the external potential \eref{extpot}.  Observe that this term is non-singular for all $x \in [-1,1]$, since $T>1$. Thus, we indeed expect that the presentation of the density \eref{ansatzdensity} of \cite{BasorChen} works here (as long as $\alpha, \beta>0$).}
\bea
\sig(x)=\frac{\sqx}{2\pi^2}\int_{a}^{b}\frac{\tv'(x)-\tv'(y)}{x-y}\:\frac{\ud y}{\sqy}~, \label{ansatzdensity}
\eea
where the limits $a$ and $b$ satisfy 
\bea
\int_{a}^{b}\frac{x\:\tv'(x)}{\sqrt{(b-x)(x-a)}}\:\frac{\ud x}{2\pi}=1~,
\;\;\;\;\;\int_{a}^{b}\frac{\tv'(x)}{\sqrt{(b-x)(x-a)}}\:\frac{\ud x}{2\pi}=0~. \label{upperlower}
\eea
Let us state some comments on the Ansatz \eref{ansatzdensity}:
\ben
\item The density \eref{ansatzdensity} is supported on the interval $(a,b)$ such that $-1<a < b <1$.  They are determined by the degree six equations described subsequently in Section \ref{sec:limits}.  There, we explicitly show that $a$ and $b$ are functions of $n$.
\item As discussed in \cite{BasorChen}, the solution \eref{ansatzdensity} can be analytically continued to negative values of $\alpha$ and $\beta$, including the values $\alpha=-1/2$ or $\beta=-1/2$ required for the $B_n$ and $D_n$ gauge groups.\footnote{The analytic continuation works up to high-order corrections in the Veneziano limit. It would also be possible to start with \emph{negative} $\alpha$ and $\beta$, and analytically continue the result to the positive values. In this case the Ans\"atze \eref{intsiggapless} and \eref{intsiggapped} in Section \ref{sec:pertanalysis} would be relevant.}
\een

\subsubsection*{Integrating \eref{ansatzdensity} to find $\sigma(x)$ explicitly}

Using \eref{extpot} we obtain
\bea \label{intgdsig}
\frac{\tv'(x)-\tv'(y)}{x-y}= -\frac{\CR}{(T-x)(T-y)} + \frac{1}{n} \left[ \frac{\alpha}{(1-x)(1-y)} + \frac{\beta}{(1+x)(1+y)} \right]~,
\eea
where $\alpha$ and $\beta$ depend on the group $G$ according to \eref{alphabeta}.
Then,
\bea
\sig(x)&=\frac{\sqx}{2\pi^2}
\Bigg[-\frac{\CR}{x-T}\int_{a}^{b}\frac{\ud y}{(y-T)\sqy}  \nn \\
& \quad + \frac{1}{n} \int_{a}^{b} \left \{ \frac{\alpha}{(1-x)(1-y)} + \frac{\beta}{(1+x)(1+y)} \right \} \frac{\ud y}{\sqy} \Bigg]~.
\eea
Using the integral \eref{intCndens}, we find that the density $\sigma(x)$ is given by
\bea
\frac{2\pi\sig(x)}{\sqrt{(b-x)(x-a)}} &=
\frac{1}{n}\left[\frac{\alpha}{(1-x)\sqrt{(1-a)(1-b)}} +\frac{\beta}{(1+x)\sqrt{(1+a)(1+b)}}
\right] \nn \\
& \qquad -\frac{\CR}{(T-x)\sqrt{(T-a)(T-b)}}~, \qquad x \in (a,b)~. \label{explicitsigma}
\eea

Let us make a few comments on \eref{explicitsigma}:
\ben
\item The terms in the first line of \eref{explicitsigma} are equal to the density computed in \cite{BasorChen}, which correspond to our analysis with $\CR=0$.  The contribution from a non-zero $\CR$ comes in as the second line of \eref{explicitsigma}.  Observe that the latter does not depend on $\alpha$ and $\beta$.
\item Recall that we take first $\alpha, \beta > 0$.  Since $-1<a<x<b<1$ and $T>1$, the second term of \eref{explicitsigma} is $\emph{strictly negative}$, whereas the first term (coming from the Jacobi weight) is \emph{strictly positive}. In order to obtain a positive charge density in the large $n$ limit, it is crucial that 1) the first term cannot be neglected and 2) the limits $a$ and $b$ must also be certain functions of $n$ in such a way that the terms in the first line can balance that in the second line.  We compute $a$ and $b$ perturbatively in the subsequent section; see Eqs. \eref{gaplessansatz} and \eref{ansatzeab}.
\item As pointed out in \cite{BasorChen}, the partition function $\CD_G(n, N_f)$ computed using \eref{explicitsigma} can be continued analytically to negative $\alpha$ and $\beta$, due to the real analyticity of the expression.   For convenience, we shall henceforth \emph{assume that $\alpha, \beta > 0$ and, at the end, we analytically continue the result to all cases listed in \eref{alphabeta}.}
\een

\subsection{The integration limits in the Coulomb gas}  \label{sec:limits}
Let us determine the constraints on the integration limits $a$ and $b$.
The first equation of \eref{upperlower} reads
\bea
 2 \pi  &= \CR \int_a^b \frac{x \: \ud x}{(x-T) \sqrt{(b-x)(x-a)}}  - \frac{\alpha}{n}  \int_a^b \frac{x\:  \ud x}{(x-1) \sqrt{(b-x)(x-a)}} \nn \\
 & \qquad  - \frac{\beta}{n}  \int_a^b \frac{x \: \ud x}{(x+1) \sqrt{(b-x)(x-a)}} ~.
\eea
Performing partial fractions and using identity \eref{intCndens}, we obtain
\bea
 2+\frac{\alpha+\beta}{n}-\CR+\frac{\CR T}{\sqrt{(T-a)(T-b)}}=\frac{1}{n}\left[\frac{\alpha}{\sqrt{(1-a)(1-b)}}+
\frac{\beta}{\sqrt{(1+a)(1+b)}}\right]~. \label{limit1}
\eea
The second equation of \eref{upperlower} reads
\bea
 0 &= \CR \int_a^b \frac{\ud x}{(x-T) \sqrt{(b-x)(x-a)}}  - \frac{\alpha}{n}  \int_a^b \frac{\ud x}{(x-1) \sqrt{(b-x)(x-a)}} \nn \\
 & \qquad - \frac{\beta}{n}  \int_a^b \frac{ \ud x}{(x+1) \sqrt{(b-x)(x-a)}}~.
\eea
Using identity \eref{intCndens}, we obtain
\bea
\frac{\CR}{\sqrt{(T-a)(T-b)}}=\frac{1}{n}\left[\frac{\alpha}{\sqrt{(1-a)(1-b)}}-
\frac{\beta}{\sqrt{(1+a)(1+b)}}\right]~. \label{limit2}
\eea
Substituting \eref{limit2} into \eref{limit1}, we obtain
\bea
2+\frac{\alpha+\beta}{n}-\CR=\frac{1}{n}\left[\frac{\alpha(1-T)}{\sqrt{(1-a)(1-b)}}+\frac{\beta(1+T)}{\sqrt{(1+a)(1+b)}}\right]. \label{limit3}
\eea
In subsequent computations, we make use of \eref{limit2} and \eref{limit3} as two constraints on the integration limits.  Note that these constraints are indeed the generalisation of the result in \cite{BasorChen} to the case of non-zero $\CR$.

\paragraph{Change of variables} 
To proceed further, let
\bea
X:=\frac{1}{\sqrt{(1-a)(1-b)}}~, \qquad Y:=\frac{1}{\sqrt{(1+a)(1+b)}}\ . \label{defXY}
\eea
and express $ab$ and $a+b$ in terms of $X$ and $Y$,
\bea
ab =\frac{1}{2}\left(\frac{1}{X^2}+\frac{1}{Y^2}\right)-1~, \qquad
a+b =\frac{1}{2}\left(\frac{1}{Y^2}-\frac{1}{X^2}\right)~. \label{abXY}
\eea
Using \eref{limit2} and \eref{abXY}, we obtain
\bea \label{XYeq1}
T^2-\frac{1}{2}\left(\frac{1}{Y^2}-\frac{1}{X^2}\right)T+\frac{1}{2}\left(\frac{1}{X^2}+\frac{1}{Y^2}\right)-1
=\frac{n^2 \CR^2}{(\alpha X- \beta Y)^2}~.
\eea
From \eref{limit3} and \eref{defXY}, we also have a $\emph{linear}$ relation between $X$ and $Y$:
\bea \label{XYeq2}
-\alpha (T-1)X+ \beta (T+1)Y=-n (\CR-2) +( \alpha+\beta)~.
\eea
We shall make use of \eref{XYeq1} and \eref{XYeq2} in subsequent computations.

\subsubsection{The degree six equation} 
Solving for $X$ from \eref{XYeq2} and substituting it into \eref{XYeq1}, we find that, upon disregarding the solution $T=1$, $Y$ satisfies the following degree six equation:
\bea
0 = \sum_{k=0}^6 c_k Y^k~,  \label{degsixeq}
\eea
where the coefficients $c_k$ are given as follows:
\bea
c_6 &= 8 \beta ^4 \widetilde{T}^3~, \nn \\
c_5 &= -8 \beta ^3 \widetilde{T}^2 \left(\alpha +\beta -n \widetilde{\CR}\right)  \left(2+\widetilde{T}\right)~, \nn \\
c_4 &=  2 \beta ^2 \widetilde{T} \Big[4 \left(\beta -n \widetilde{\CR}\right) \left(2 \alpha +\beta -n \widetilde{\CR}\right)+2 \widetilde{T} \Big \{ 4 n^2+(\alpha +\beta ) (5 \alpha +3 \beta ) \nn \\
& \quad +n \widetilde{\CR} \left(4 (n-2 (\alpha +\beta ))+5 n \widetilde{\CR}\right)\Big\} +(2 n+\alpha +\beta ) \left(-2 n+\alpha +\beta -2 n \widetilde{\CR}\right) \widetilde{T}^2 \Big]~, \nn \\
c_3 &= -8 \beta \widetilde{T}  \left(\alpha +\beta -n \widetilde{\CR}\right) \Big[ 4 n^2+2 \alpha  \beta +2 n \widetilde{\CR} \left(2 n-\alpha -\beta +n \widetilde{\CR}\right) \nn \\
& \quad + \left(-2 n^2+\alpha  (\alpha +\beta )-n (2 n+\alpha +\beta ) \widetilde{\CR}\right) \widetilde{T}\Big]~, \nn \\
c_2 &= \left(\alpha +\beta -n \widetilde{\CR}\right)^2 \Big[ 16 n^2-4 \beta ^2+4 n^2 \widetilde{\CR} \left(4+\widetilde{\CR}\right)-2  \widetilde{T}\Big\{4 n^2+\beta  (-2 \alpha +3 \beta ) \nn \\
& \quad +2 n (2 n+\alpha +\beta ) \widetilde{\CR}\Big \} +(\alpha -\beta ) (\alpha +\beta ) \widetilde{T}^2 \Big]~, \nn \\
c_1 &= 2 \beta  \left(\alpha +\beta -n \widetilde{\CR}\right)^3 \left(2+\widetilde{T}\right)~, \nn \\
c_0 &= -\left(\alpha +\beta -n \widetilde{\CR}\right)^4~,
\eea
with
\bea
\widetilde{\CR} = \CR-2~,\qquad \widetilde{T} = T+1~.
\eea
One can also similarly compute the degree six equation for $X$.

We conjecture that these degree six equations arise from the existence of the Painlev\'e VI equation that the Hilbert series satisfies \cite{Basor:2011da}.

\subsection{The free energy}  \label{sec:freeenergy}
Recall from \eref{freeenA} that the free energy receives two contributions, from $A$ and $I$.

\subsubsection*{Computing $A$ from \eref{constantA}} 
We compute $A$ as follows.  First, let us multiply both sides of \eref{constantA} by the factor $\frac{1}{\pi} [(b-x)(x-a)]^{-1/2}$ and integrate over $x$:
\bea
\frac{A}{\pi} \int_{a}^b \frac{\ud x}{\sqrt{(b-x)(x-a)}} &= \int_{a}^b\frac{\sv(x)}{\pi \sqrt{(b-x)(x-a)}}  \ud x \nn \\
& \qquad - 2 \int_{a}^b \frac{\ud x}{\pi \sqrt{(b-x)(x-a)}} \int_{a}^b \log |x-y| \sig(y)  \ud y~,
\eea
and so
\bea
A   = \int_{a}^b\frac{\sv(x)}{\pi \sqrt{(b-x)(x-a)}}  \ud x - 2 \int_{a}^b \ud y ~ \sigma(y) \int_a^b \frac{\log |x-y|}{\pi \sqrt{(b-x)(x-a)}} \ud x \label{Aint}
\eea
Note the following important observation (see, \eg, (6.19) of \cite{ChenMcKay}):
\bea
\frac{\partial}{\partial y}  \int_a^b \frac{\log |x-y|}{ \sqrt{(b-x)(x-a)}} \ud x = \CP \int_a^b \frac{\ud x }{(y-x)  \sqrt{(b-x)(x-a)}} =0~,
\eea
if $y \in [a,b]$.
In other words, the integral with respect to $x$ in the second term on the RHS of \eref{Aint} is independent of $y$.  Hence, we can replace $y$ in the logarithm by a constant, say $a$:
\bea
\int_a^b \frac{\log |x-y|}{ \sqrt{(b-x)(x-a)}} \ud x = \int_a^b \frac{\log |x-a|}{ \sqrt{(b-x)(x-a)}} \ud x~.
\eea
Therefore,
\bea
A  &= \int_{a}^b\frac{\sv(x)}{\pi \sqrt{(b-x)(x-a)}}  \ud x -2 \left( \int_{a}^b \ud y ~ \sigma(y) \right) \left( \int_a^b \ud x \frac{\log (x-a)}{\pi \sqrt{(b-x)(x-a)}}  \right) \nn \\
&= \int_{a}^b\frac{\sv(x)}{\pi \sqrt{(b-x)(x-a)}} \ud x - 2 \int_a^b \frac{\log (x-a)}{\pi \sqrt{(b-x)(x-a)}} \ud x  \nn \\
&= \int_{a}^b\frac{\sv(x)}{\pi \sqrt{(b-x)(x-a)}} \ud x - 4 \log \left(\frac{1}{2} \sqrt{b-a} \right)~, \label{Aexact}
\eea
Observe that this expression of $A$ is independent of the charge density $\sigma(x)$.

Using \eref{extpot} and \eref{integralA2}, the integral in the first term is
\bea
\int_{a}^b\frac{\sv(x)}{\pi \sqrt{(b-x)(x-a)}} \ud x  & = 2\CR \log \left( \frac{\sqrt{T-a} + \sqrt{T-b}}{2} \right) \nn \\
& \quad - \frac{2}{n} \left[ \alpha \log \left(\frac{\sqrt{1-a} + \sqrt{1-b}}{2}  \right)+\beta \log \left( \frac{\sqrt{1+a} + \sqrt{1+b}}{2}  \right) \right] ~.
\eea
Thus, we obtain
\bea \label{Ares}
A &=2\CR \log \left( \frac{\sqrt{T-a} + \sqrt{T-b}}{2} \right) - 4 \log \left(\frac{1}{2} \sqrt{b-a} \right) \nn \\
& \quad - \frac{2}{n} \left[ \alpha \log \left( \frac{\sqrt{1-a} + \sqrt{1-b}}{2}  \right) + \beta \log\left( \frac{\sqrt{1+a} + \sqrt{1+b}}{2}  \right) \right]~.
\eea

\paragraph{Computing the integral $I$ in \eref{freeenc}}
\bea \label{Idef}
I &= \int_a^b \ud x ~{\sf v}(x) \sigma(x) \nn \\
&=   \CR \int_{a}^b \ud x\;\log(T-x) \sigma(x) - \frac{1}{n} \int_a^b \ud x  \sigma(x) \left[ \alpha \log (1-x) +\beta \log (1+x)\right]~.
\eea
Using \eref{explicitsigma}, we have
{\small
\bea
I &= \frac{1}{2 \pi n}\Bigg[ \frac{\alpha \CR}{\sqrt{(1-a)(1-b)}} I^-_1  + \frac{\beta \CR}{\sqrt{(1+a)(1+b)}} I^+_1 -\frac{\alpha^2}{n\sqrt{(1-a)(1-b)}} I^{--}_2 - \frac{\beta^2}{n\sqrt{(1+a)(1+b)}} I^{++}_2   \nn \\
& \qquad \qquad -\frac{\alpha \beta}{n\sqrt{(1-a)(1-b)}} I^{+-}_2 - \frac{\alpha \beta}{n\sqrt{(1+a)(1+b)}} I^{-+}_2 \Bigg]  \nn \\
& \qquad \qquad -\frac{\CR}{2 \pi \sqrt{(T-a)(T-b)}} \Bigg[ I_3  \CR  - \frac{\alpha}{n} I^{-}_4 - \frac{\beta}{n} I^{+}_4 \Bigg]~,
\eea}
where the integrals $I^\pm_1, I^{\pm \pm}_2, I_3, I^{\pm}_4$ are defined as
\bea
\begin{array}{lll}
I^\pm_1 &= \int_a^b \ud x \frac{\log (T-x) \sqrt{(b-x)(x-a)} }{(1 \pm x)}~,  \qquad I^{- \pm}_2 &=  \int_a^b \ud x \frac{ \log (1- x) \sqrt{(b-x)(x-a)}}{1 \pm x}  \\
I^{+ \pm}_2 &=  \int_a^b \ud x \frac{\log (1+ x)  \sqrt{(b-x)(x-a)}}{1 \pm x}~,  \qquad I_3 &=  \int_a^b \ud x \frac{\log (T-x) \sqrt{(b-x)(x-a)} }{(T-x)}  \\
I^{\pm}_4 &=  \int_a^b \ud x  \frac{\log (1 \pm x) \sqrt{(b-x)(x-a)}}{T-x}~;
\end{array}
\eea
they are computed explicitly in Appendix \ref{app:intF}.

\section{Series expansions in the Veneziano limit}
\label{sec:pertsolutions}
In this section the main goal is to compute, from the preceding exact results, the charge density, the free energy, and hence the Hilbert series perturbatively in the Veneziano limit, i.e., as a power series in $1/n$ while keeping the ratio $\CR$ defined in \eref{defR} fixed and finite.  

As it turns out (see also the analysis of Section~\ref{sec:pertanalysis}), the result is essentially different depending on whether the system is in one of the following phases:
\ben
\item {\bf The gapless phase:}  The name ``gapless'' denotes the phase in which the integration limits $a$ and $b$ are such that $a \rightarrow -1, b \rightarrow 1$ as $n \rightarrow \infty$.  In this phase, the charge density is a continuous function supported on the interval $(-1,1)$ in the large $n$ limit.  In other words, there is {\it no gap} created on the charged interval even in the presence of external interactions.  
\item {\bf The gapped phase:} On the other hand, in the ``gapped phase'', the lower limit $a$ tends to $a_0$, for some $a_0$ strictly greater than $-1$, as $n \rightarrow \infty$. In the latter phase, the gap is created in the interval $(-1, a_0)$ and the charge density is supported only on the interval $(a_0,1)$ in the large $n$ limit.  The presence of such a gap is due to the attractive interaction due to the external particle at $x=T$. 
\een

Subsequently, we show that the relevant parameters involved in these two phases are $\CR$ {\it and} $T$. Explicitly, we have the following:
\bi
\item {\bf The gapless phase:} $0\leq \CR \leq 2$, \emph{or} $\CR >2$ and $T \geq T_c$ (\ie, $0< t\leq t_c$)~,
\item {\bf The gapped phase:} $\CR > 2$ and $1 < T < T_c$ (\ie, $t_c < t< 1$)~,
\ei
where the relation between $T$ and $t$ is given in \eref{Tandt} and
\bea
t_c &= \frac{1}{\CR-1}~, \qquad T_c = \frac{\CR(\CR-2)+2}{2(\CR-1)}~. \label{criticalT}
\eea
Notice that taking $\CR = 2 N_f/n$ for the gauge groups $U(n)$ and $SU(n)$, the critical value $t_c$ agrees with that of \cite{Jokela:2011vg}.

We emphasise again that in the following computation $\alpha, \beta > 0$ is assumed; the end result is then analytically continued also to the cases $\alpha, \beta = -1/2$.

\subsection{The gapless phase} \label{sec:gaplesssoln}
The first task is to find the endpoints $a$ and $b$ perturbatively at large $n$. In this phase, we require that $a \rightarrow -1$ and $b \rightarrow 1$ as $n \rightarrow \infty$. We therefore propose the following Ans\"atze: 
\bea \label{gaplessansatz}
 a = -1 + \frac{a_2}{n^2}+ \frac{a_3}{n^3} + \morder{\frac{1}{n^4}}~,\qquad b = 1 - \frac{b_2}{n^2}- \frac{b_3}{n^3} + \morder{\frac{1}{n^4}}~,
\eea
where $a_2, a_3, b_2, b_3$ are fixed and finite such that $a_2, b_2 >0$, so that the interval $[a,b]$ is contained in $[-1,1]$ as $n \rightarrow \infty$.

Inserting these Ans\"atze in~\eqref{XYeq1} and~\eqref{XYeq2} and recalling the definitions \eref{defXY}, we find the perturbative solution
\bea
\begin{array}{ll}
a_2  = \frac{2 (T+1)^2 \beta ^2}{\left[(2-R)(T+1)-R \sqrt{T^2-1}\right]^2}~, &  \qquad
b_2  = \frac{2 (T-1)^2 \alpha ^2}{\left[(2-R)(T-1)+R \sqrt{T^2-1}\right]^2}~,  \\
a_3  = \frac{4 (T+1)^3 \beta ^2 (\alpha +\beta)}{\left[(2-R)(T+1)-R \sqrt{T^2-1}\right]^3}~, &  \qquad
b_3  = -\frac{4 (T-1)^3 \alpha ^2 (\alpha +\beta )}{\left[(2-R)(T-1)+R \sqrt{T^2-1}\right]^3}~.
\end{array}
\eea
Observe that in this solution, $a_2$ and $b_2$ are strictly positive for $T>1$, as required.

\subsubsection{The charge density} 
Substituting \eqref{gaplessansatz} into \eqref{explicitsigma}, we obtain
\bea
 \sigma(x) &= \frac{(2-\CR) (T-x)+\CR \sqrt{T^2-1}}{2 \pi  (T-x) \sqrt{1-x^2}} +\frac{1}{n} \left(\frac{\alpha +\beta }{2 \pi  \sqrt{1-x^2}}\right)+ \morder{\frac{1}{n^2}} \label{densitygapless}
\eea
for all $x \in (-1,1)$.
We analytically continue this result to all cases listed in \eref{alphabeta}.  

Observe that the leading contribution does not depend on $\alpha$ and $\beta$.  It is non-negative for all $x \in [-1,1]$ if and only if 
\bea
0 \leq \CR \leq 2 \quad \text{\it or} \quad \CR>2~\text{and}~T \geq T_c :=\frac{\CR(\CR-2)+2}{2(\CR-1)}~.
\eea
When the subleading contribution is included, we consider each case as follows.
\bi
\item For $G=C_n$, $\alpha=\beta=1/2$ and so the subleading term is strictly positive.\footnote{This statement is true for all $x$ within the open interval $(-1,1)$. The expansion is involved at the endpoints, which is reflected by the fact that~\eqref{densitygapless} does not satisfy the normalization condition at next-to-leading order.}
\item For $G=B_n$, $\alpha=-\beta=1/2$ and so the subleading term as written in \eref{densitygapless} vanishes.  Formally, the subdominant contribution suppressed by $1/n^2$ can be computed by expanding $a$ and $b$ further to include terms at order $\morder{n^{-4}}$.  However, the result for the latter is too lengthy to be reported here, and it is not likely that the result at such an order will be useful, because the relation~\eqref{FandD} is expected to receive nontrivial correction at high orders.
\item For $G=D_n$, $\alpha=\beta = -1/2$ and so the subleading term, suppressed by $1/n$, is strictly negative.   

One can ask when the perturbative density \eref{densitygapless} breaks down; in other words, when $\sigma(x)$ contains a zero for some $x \in [-1,1]$.
The zero of $\sigma(x)$ occurs at
\bea
x_c = \left[ T+\frac{\CR}{2-\CR}\sqrt{T^2-1} \right] + \frac{1}{n} \left[ \frac{\CR \sqrt{T^2-1}}{(2-\CR)^2} \right]~.
\eea
For $0 \leq \CR \leq 2$, it is immediate that $x_c >1$, which lies outside the support $[-1,1]$ of the density; thus, $\sigma(x)$ is strictly positive for all $x \in [a,b]$.  On the other hand, for $\CR >2$ and $T> T_c$, $x_c \in [-1,1]$ and the charge density thus becomes negative if and only if
\bea
T_c< T<\frac{2\CR(\CR-1)+1}{2\CR-1} \quad \text{and} \quad 1 < n \leq \frac{(\CR-2)(1+T) + \CR\sqrt{T^2-1}}{2[2(\CR-1)(T+1)-\CR^2]}~.
\eea
Since $n$ is bounded from above by a finite number, higher orders in the perturbation theory become significant.
\ei
These arguments support the validity of analytic continuation we performed earlier.

\subsubsection{The free energy} 
Let us now compute the free energy $F$ given by \eref{freeenA}.   There are two contributions, from the constant $A$ and the integral $I$.
 
\paragraph{The contribution from $A$}
The constant $A$ of~\eqref{Ares} is
\bea
A &= \left[ 2\log 2 -\CR \log(2t) \right]+\frac{1}{n}(\alpha+\beta)\log 2 + \frac{1}{n^2} \Delta^{(-2)}_A  + \morder{n^{-3}} ~. \label{Agapless}
\eea
where $t$ and $T$ are related by \eref{Tandt} and
\bea
\Delta^{(-2)}_A &= \frac{1}{2} \left[ - \frac{(1-t) \alpha ^2}{(\CR-1) t+1}+\frac{(1+t) \beta ^2}{(\CR-1) t-1} \right]~.
\eea
It should be observed from \eref{Agapless} that the leading term does not depend on $\alpha$ and $\beta$ and is hence universal for all groups listed in \eref{alphabeta}. 

\paragraph{The contribution from $I$} The integral $I$ is defined in~\eqref{Idef}.  A direct computation yields
\bea
I &= \CR \left[\CR \log\left(1-t^2\right)- \log (2t)\right] + \frac{1}{n} \Big[ (\alpha +\beta ) \log2 \nn \\
& \qquad  -2 \CR \{\alpha \log (1-t)+\beta \log(1+t) \} \Big]  + \morder{n^{-2}}~. \label{Igapless}
\eea

\paragraph{The free energy}  By inserting these expressions in~\eqref{freeenA} the free energy becomes
\bea
 F_{\text{gapless}} &= n^2 \left[ \log 2 -\CR \log(2t) + \frac{1}{2} \CR^2 \log(1-t^2) \right]  +  n \Big[ (\alpha +\beta ) \log 2 \nn \\
 & \qquad \qquad -\CR \{\alpha  \log(1-t)+\beta  \log(1+t)\} \Big]+ \morder{{n^0}}~. \label{Fgapless}
\eea
Again, we emphasise that the leading contribution does not depend on $\alpha$ and $\beta$ and is hence universal for all groups listed in \eref{alphabeta}. 

\subsubsection{The Hilbert series} \label{sec:HSgapless}
We can now readily compute the partition functions and hence the Hilbert series from \eqref{FandD}.
\bea
\log \CD_G \sim -F_{\text{gapless}}&= -n^2 \left[ \log 2 -\CR \log(2t) + \frac{1}{2} \CR^2 \log(1-t^2) \right]  -  n \Big[ (\alpha +\beta ) \log 2 \nn \\
 & \qquad \qquad -\CR \{\alpha  \log(1-t)+\beta  \log(1+t)\} \Big]+ \morder{{n^0}}~.
\eea
Using \eref{alphabeta}, \eref{defR} and \eref{HS}, we find that the Hilbert series are given by
\bea \label{HSgapless}
\log g_{N_f, G}(t) &\sim \begin{cases}
-\frac{1}{2}N_f(N_f+1) \log(1-t^2)~, &\qquad G=B_n,\; D_n \\
-(2N_f-1)N_f \log(1-t^2)~, &\qquad G=C_n~.
\end{cases}
\eea
Observe that these asymptotic results turn out to coincide with the exact results given by \cite{Hanany:2008kn,Basor:2011da} for the case $0\leq \CR<2$.  In addition, the negative coefficient of $ \log(1-t^2)$ indeed coincides with the dimension of the moduli space for any $\CR \geq 0$.

\subsection{The gapped phase}  \label{sec:gappedphase}
In the gapped phase, we take Ans\"atze for $a$ and $b$ to be
\bea \label{ansatzeab}
 a = a_0 + \frac{a_1}{n} + \morder{\frac{1}{n^2}}~,\qquad b = 1 - \frac{b_2}{n^2}- \frac{b_3}{n^3} + \morder{\frac{1}{n^4}}~,
 \eea
where $a_0, a_1, b_2, b_3$ are fixed such that $a_0 >-1$ and $b_2>0$, so that the interval $[a,b]$ is contained in $[-1,1]$ as $n \rightarrow \infty$.
 
These coefficients are determined by substituting these Ans\"atze to \eref{XYeq1} and \eref{XYeq2}.  In order to calculate the charge density and free energy at leading order, only the coefficients
\bea \label{a0b2}
 a_0 = \frac{\CR^2-4 (\CR-1) T}{(\CR-2)^2}~,\qquad b_2 =\frac{(T-1)\alpha^2}{4 (\CR-1)}
\eea
are needed.  Notice that $a_0$ is the same as in~\eqref{eq:asol}.  For the subleading order, it is necessary to take the following coefficients into account:
\bea
a_1 &= -\frac{2 \CR^2 (T-1) [S (\alpha +\beta )-(\CR-2) \beta ]}{(\CR-2)^3 S}~, \label{eq:a1} \\
b_3 &= -\frac{\alpha ^2(T-1) [(\CR-2) \alpha +(\CR-S-2) \beta ]}{8 (\CR-1)^2}~, \label{eq:b3}
\eea
where we define
\bea
S &= \sqrt{\CR^2-2(\CR-1) (1+T)} = \sqrt{2(\CR-1)(T_c-T)}~. \label{defS}
\eea

Note that the quantity in the square root in \eref{defS} is non-negative, $a_0 >-1$ and $b_2>0$ if and only if
\bea \label{cond1gapped}
(1 < T < T_c \quad \text{and} \quad \CR > 2) \qquad \text{or} \qquad (1 < T < T_c \quad \text{and} \quad 1<\CR <2).
\eea
In the following, we use the positivity of the charge density to rule out the second condition.

\subsubsection{The charge density} 
Using the above coefficients, we can compute the density from \eqref{explicitsigma} which yields
\bea
 \sigma(x) 
 &=  \frac{(\CR-2)\sqrt{x-a_0}}{2\pi(T-x)\sqrt{1-x}} + \frac{1}{n} \Delta^{(-1)}_\sigma  (x)+ \morder{n^{-2}}\ , \label{densitygapped}
\eea
where $a_0$ is given in \eref{a0b2} and the subleading term $\Delta^{(-1)}_\sigma  (x)$, suppressed by $1/n$, is given by
\bea
\Delta^{(-1)}_\sigma  (x)= \frac{ (\CR-2) (1+x) (\alpha +\beta )-2 S \beta}{2 \pi  (\CR-2)  (1+x) \sqrt{(x-a_0) (1-x)}}~. \label{subleadingdensity}
\eea

Observe that the leading order of $\sigma(x)$ is positive definite if the first condition in \eref{cond1gapped} holds.  Hence, the condition for the gapped phase is
\bea
1 < T < T_c \quad \text{and} \quad \CR > 2~.
\eea
We shall henceforth assume this condition throughout this section.

Note that the charge density as given in \eref{densitygapped} can be analytically continued to all cases listed in \eref{alphabeta}.

\subsubsection{The free energy}  
Let us now compute the free energy.   There are two contributions from the constant $A$ and the integral $I$.

The potential term $A$ of~\eqref{Ares} is
\bea \label{Agapped}
  A &= 2(\CR-1) \log (\CR-1)-2(\CR-2) \log(\CR-2)+(\CR-2)\log  (T-1)  \nn \\
  & \quad + \frac{1}{n} \Delta^{(-1)}_A+ \morder{\frac{1}{n}}~,
\eea
and the subleading term, suppressed by $1/n$, is
\bea
\Delta^{(-1)}_A  &=\beta  \log2+2 (\alpha +\beta ) \log(\CR-2) \nn \\
& \quad -2 \beta  \log(\CR+S-2)-\alpha  \log \{ (\CR-1) (T-1) \}~.
\eea

The integral $I$ defined in~\eqref{Idef} becomes
\bea \label{Igapped}
 I &=-\CR \left[\CR \log\left(\frac{(\CR-1)^2}{\CR(\CR-2)}\right)-2 \log\left(\frac{(\CR-1) \sqrt{T-1}}{\CR-2}\right)\right] + \frac{1}{n} \Delta^{(-1)}_I + \morder{n^{-2}}~,
\eea
and the subleading term $\Delta^{(-1)}_I $ is
\bea
\Delta^{(-1)}_I  &= -(\CR-1) \beta  \log 2 -2 (\CR-1) (\alpha +\beta ) \log (\CR-2)+2 \CR (\alpha +\beta ) \log (\CR-1) \nn \\
& \quad +2 (\CR-1) \beta  \log (\CR+S-2)-\alpha  \log \{(\CR-1) (T-1)\} \nn \\
& \quad -\CR \beta  \log \{ 1+\CR (\CR+S-T-1)+T \}~.
\eea

Substituting the expressions for $A$ and $I$ in~\eqref{freeenA}, the free energy is
\bea \label{finalfreeen}
F_{\text{gapped}} 
&= \frac{1}{2} n^2 \Big[ \CR^2 \log \{ \CR(\CR-2) \}+2 \CR \log (T-1) -2 \log \{ (\CR-1)(T-1) \} \nn \\
& \qquad \quad -4 (\CR-1) \log (\CR-2) -2 (\CR-2) \CR \log (\CR-1) \Big] \nn \\
& \quad - \frac{1}{2} n \Bigg[ (\CR-2)\beta \log2 +2 (\alpha +\beta )(\CR-2) \log(\CR-2) -2\CR (\alpha +\beta ) \log(\CR-1) \nn \\
& \qquad \quad -2(\CR-2)\beta \log (\CR+S-2) +2 \alpha \log\{(\CR-1) (T-1)\} \nn \\
& \qquad \quad + \CR \beta  \log\{ 1+\CR (\CR+S-T-1)+T \} \Bigg] + \morder{n^0}~.
\eea
The leading order term is again independent of $\alpha$ and $\beta$ and thus universal. Interestingly, one can check that this term (as well as the one in gapless phase,~\eqref{Fgapless}) also agrees with the one for the $U(n)$ and $SU(n)$ gauge groups (given in Eq.~(55) of~\cite{Jokela:2011vg}), after the leading order factors from the relations~\eqref{HS} are included.
We also emphasise that the free energy \eref{finalfreeen} can be analytically continued to all cases listed in \eref{alphabeta}.

\subsubsection{The Hilbert series}\label{sec:Hilb}
We use \eref{FandD} to compute the partition function and the Hilbert series follows easily from \eref{HS}.  For illustration, let us present the result in the limit $t \rightarrow 1$, where the expressions simplify; we set
\bea
t = 1- \eta~, \qquad \text{with} \quad \eta \rightarrow 0~.
\eea
We thus obtain
\bea
\log g_{N_f, C_n} &= n \log(2)+\Big[\frac{1}{2} (\CR-2) n +2 (\CR-1) n^2\Big] \log(\CR-2) \nn \\
& \quad +\Big[-\frac{1}{2} (\CR-1) n+(\CR-1)^2 n^2\Big] \log(\CR-1) \nn \\
& \quad -\frac{1}{2} \CR^2 n^2 \log \{ (\CR-2) \CR\}-\Big[2(\CR-1)n^2 -n \Big] \log(\eta ) + \morder{\eta ; n^0}~, \label{HSCngap} \\
\log g_{N_f, B_n} &= - n \log(2)+\Big[\frac{1}{2} (\CR-2) n+2 (\CR-1) n^2 \Big] \log(\CR-2) \nn \\
& \quad +\Big[-\frac{1}{2} (\CR-1) n+(\CR-1)^2 n^2\Big] \log(\CR-1) \nn \\
& \quad -\frac{1}{2} \CR^2 n^2 \log \{ (\CR-2) \CR \}-\Big[n(\CR-1) (2 n+1)\Big] \log(\eta ) + \morder{\eta ; n^0} ~, \label{HSBngap} \\
\log g_{N_f,D_n} &= -n \log(2)+\Big[-\frac{1}{2} (\CR-2) n+2 (\CR-1) n^2\Big] \log(\CR-2) \nn \\
& \quad +\Big[\frac{1}{2} (\CR-1) n+(\CR-1)^2 n^2\Big]  \log(\CR-1) \nn \\
& \quad -\frac{1}{2} \CR^2 n^2 \log \{ (\CR-2) \CR \}-\Big[2(\CR-1)n^2 +n \Big] \log(\eta ) + \morder{\eta ; n^0} ~. \label{HSDngap}
\eea
Note that the minus of the coefficient of $\log \eta = \log(1-t)$ in each case indeed coincides with the dimension of the moduli space as given in \cite{Hanany:2008kn,Basor:2011da}.

Moreover, we can perform further non-trivial checks of these expressions by considering the following examples:
\bi
\item {\bf $G=C_n$ with $\CR = 2+2n^{-1}$ or $N_f = n+1$.} Eqs. \eref{HSCngap} and (4.27) of \cite{Basor:2011da} are in agreement with the limits $\eta \rightarrow 0,\; n \rightarrow \infty$:
\bea
\log g_{N_f, C_n} &= \log(1-t^{2N_f}) - N_f(2N_f-1) \log(1-t^2) \nn \\
&\sim -n (3+2 n) \log (2 \eta)~. \qquad  
\eea
\item {\bf $G=B_n$ with $\CR = 2+n^{-1}$ or $N_f = 2n+1$.} Eqs. \eref{HSBngap} and (2.36) of \cite{Basor:2011da} are in agreement with the limits $\eta \rightarrow 0,\; n \rightarrow \infty$:
\bea
\log g_{N_f, B_n} &= \log(1+t^{N_f}) - \frac{1}{2}N_f(N_f+1) \log(1-t^2) \nn \\
&\sim -n (3+2 n) \log (2 \eta)~. \qquad  
\eea
\item {\bf $G=D_n$ with $\CR = 2+n^{-1}$ or $N_f = 2n+1$.} Eqs. \eref{HSDngap} and (3.31) of \cite{Basor:2011da} are in agreement with the limits $\eta \rightarrow 0,\; n \rightarrow \infty$:
\bea
\log g_{N_f, B_n} &= \log[1+(1+2n)t^{2n}-(1+2n)t^{2n+2}-t^{4n+2}] - \frac{1}{2}N_f(N_f+1) \log(1-t^2) \nn \\
&\sim -n (3+2 n) \log (2 \eta)~. \qquad  
\eea
\ei

\section{Leading order approach in the Veneziano limit} \label{sec:pertanalysis}
In this section, we present an alternative approach to derive most of the results to the Hilbert series presented above. The purpose is to bypass the exact results presented in Section \ref{sec:exactv} and inferring the results more simply by using standard perturbation theory; see \cite{Jokela:2010cc,Jokela:2010zp,Jokela:2009fd,Mandal:1989ry}. Solving the equilibrium condition at leading order (LO) in the Veneziano limit is enough to obtain the free energy and hence the Hilbert series up to next-to-leading order (NLO), as we will demonstrate below.

To be precise, we write the potential as 
\bea
 \tv(x) = \tv_0(x) + \frac{1}{n} \Delta_\tv^{(-1)}(x)~,
\eea
where
\bea \label{LOv}
 \tv_0(x) =  \CR \log(T-x) 
\eea
is the leading contribution in the Veneziano limit, and the contribution from the Jacobi weight
\bea \label{NLOv}
 \Delta_\tv^{(-1)}(x) = - \log w_G = - \alpha \log(1-x) - \beta \log (1+x) 
\eea
appears at NLO. Then, generalizing the notation of the previous sections, we can expand the charge density
\bea
 \sigma(x) = \sigma_0(x) + \frac{1}{n} \Delta_\sigma^{(-1)}(x) + \frac{1}{n^2} \Delta_\sigma^{(-2)}(x) + \cdots \ .
\eea
Below we will use analogous notation for series coefficients for other quantities like $A$ and $I$.
The terms in this expansion can then be solved by applying standard perturbation theory to the equilibrium condition~\eqref{constantA}. However, as we already mentioned above, for our purposes it is enough to restrict to the solution at LO.\footnote{Finding the perturbative solutions at high orders is involved, in particular when $\alpha$ and $\beta$ are positive, due to singular behavior near the endpoints of integration. At our LO analysis such complications are absent.} As is usual in perturbation theory, the NLO correction to free energy can be obtained in terms of the LO solution and the NLO ``source'' of perturbation, i.e, the Jacobi weight of~\eqref{NLOv}.

\subsection{Leading order solution} \label{app:LO}
In the LO computation, only the LO piece~\eqref{LOv} contributes to the external potential. The equilibrium condition \eref{constantA} thus reads at LO
\bea \label{eq:poteq2}
\tv_0(x) -2 \int_{a}^b  \ud y~  \log|x -y|  \sigma_0(y)  = \text{const.} = A_0 ~,
\eea
where the potential is given in~\eqref{LOv}, and the support of $\sigma_0(x)$ is $[a,b]$.

Before going to the explicit solutions to~\eqref{eq:poteq2}, let us discuss the physical interpretation of the system. 
Recall that $\sigma(x)$ contains only carriers with charges having a single sign, and therefore it cannot become negative.  The external charge at $x=T >1$ has the opposite charge to the charge density $\sigma(x)$ on the interval $-1<x<1$, and therefore attracts it. Therefore, we expect that we will always have $b=1$. We further expect to  have two kinds of solutions to \eqref{eq:poteq2} depending on the values of $T$ and $\CR$:
\ben
\item  For a sufficiently large $T$ or sufficiently small $\CR$, the attraction is not sufficiently large to move the end point $x=-1$ of the interval and hence the lower limit $a=-1$ in this case.  Note that in this case, the charge density $\sigma_0(x)$ blows up at both end points $x= \pm1$.  
\item As $\CR$ is increased and $T$ is decreased to sufficiently large and small values, respectively, the attraction drags the end point $x=-1$ of the interval to $x=a_0 >-1$ and therefore a gap appears for $x \in [-1,a_0)$.  When the gap has appeared, the charge density must vanish continuously at $x=a_0$ (but it still blows up at $x=1$), since otherwise there would be a net force acting on the charges near the edge of the gap, so that moving the charges into the gap would lower the energy of the system.
\een
The first situation is known as the {\it gapless phase} and the second is known as the {\it gapped phase}, which were already discussed in Section~\ref{sec:pertsolutions}.
We will determine $a_0$ and the values of $\CR$ and $T$ at the transition point below.

\subsubsection{Charge density in the gapless phase}  
According to the above discussion, we expect that for the unique physical, positive definite solution to~\eqref{eq:poteq2} in the gapless phase, the density is supported in the interval $[-1,1]$ with
\bea \label{assumpgapless}
\sigma_0(x) \rightarrow +\infty \quad \text{as} \quad x \rightarrow \pm 1^\mp~.
\eea

The solution of \eref{extforce} subject to the boundary conditions \eref{assumpgapless} is\footnote{See, e.g., Eq. (26) on Page 131 of \cite{mikhlin1964integral}.}
\bea \label{intsiggapless}
\sig_0(x)&=\frac{1}{2\pi^2\sqrt{1-x^2}}\, \CP \int_{-1}^{1}\frac{\tv_0'(y)}{y-x}\:\sqrt{1-y^2}\:\ud y + \frac{c_1}{\sqrt{1-x^2}}~.
\eea
Here the constant $c_1$ can be determined from the normalisation condition \eref{normsig}. 
Note that the integral over $x$ from $-1$ to $1$ of the first term in \eref{intsiggapless} vanishes due to the integral identity \eref{intprincipal}.
We therefore obtain
\bea
c_1 = 1/ \pi~.
\eea

Substituting \eref{LOv} into \eref{intsiggapless} and using the identity \eref{intprinc2}, we find
\bea \label{sigmagpl}
\sigma_0(x) &= \frac{\CR}{2 \pi^2 \sqrt{1-x^2}} \; \CP \int_{-1}^1 \frac{\sqrt{1-y^2}}{(y-x)(y-T)} \ud y + \frac{1}{\pi \sqrt{1-x^2}} \nn \\
&= \frac{(2-\CR) (T-x)+\CR \sqrt{T^2-1}}{2 \pi  (T-x) \sqrt{1-x^2}}  ~.
\eea

\paragraph{Conditions for gap formation} 
Let us then discuss when the solution~\eqref{sigmagpl} is valid. If $\CR \le 2$, it is easy to see that $\sigma(x)$ is nonnegative for all acceptable values of $t$ or $T$. Therefore it is the desired solution. For $\CR>2$, the numerator of $\sigma$ takes its smallest value at $x=-1$. Requiring this to be nonnegative leads to
\bea
(2-\CR) (T+1)+\CR \sqrt{T^2-1} \ge 0 \qquad \text{or} \qquad \frac{1+t}{t} \ge \CR~.
\eea
This translates to
\bea
T \ge T_c = \frac{\CR(\CR-2)+2}{2(\CR-1)} \qquad \text{or} \qquad  0 \le t \le t_c = \frac{1}{\CR-1}~.
\eea

\subsubsection{Charge density in the gapped phase}
We expect that the physically reasonable solution in the gapped phase satisfies
\bea\label{assumpgapped}
\sigma_0(x) \rightarrow +\infty~ \text{as}~ x \rightarrow 1^-, \quad \text{and} \quad \sigma_0(x) \rightarrow 0~\text{as}~x \rightarrow a_0^+~.
\eea

The unique solution of \eref{extforce} subject to the boundary conditions \eref{assumpgapped} takes the form\footnote{See, {\emph{e.g.}},  Eq. (25) on Page 131 of \cite{mikhlin1964integral}.}
\bea \label{intsiggapped}
\sig_0(x)&=\frac{1}{2\pi^2}\sqrt{\frac{x-a_0}{1-x}}\, \CP \int_{a_0}^{1}\frac{\tv_0'(y)}{y-x}\:\sqrt{\frac{1-y}{y-a_0}}\:\ud y ~.
\eea
In order to compute $a_0$, we use the normalisation condition
\bea
1 &= \int_{a_0}^1 \sigma_0(x) \ud x = - \frac{1}{2 \pi} \int_{a_0}^1 \tv_0'(x) \sqrt{\frac{1-y}{y-a_0}} \ud y = \frac{\CR}{2\pi} \int_{a_0}^1 \frac{1}{T-x} \sqrt{\frac{1-y}{y-a_0}} \ud y \nn \\
&= \frac{\CR}{2} \left( 1 - \sqrt{\frac{T-1}{T-a_0} }\right)~.
\eea
Thus, we obtain
\bea
a_0 = \frac{\CR^2-4 (\CR-1) T}{(\CR-2)^2}~. \label{asol}
\eea

Substituting \eref{LOv} into \eref{intsiggapped}, we obtain the leading term of the density to be
\bea
\sig_0(x)&= \frac{\CR}{2\pi^2} \sqrt{\frac{x-a_0}{1-x}} \CP \int_{a_0}^1 \frac{1}{(y-T)(y-x)} \sqrt{\frac{1-y}{y-a_0}} \ud y  \nn \\
&= \frac{(\CR-2) \sqrt{x-a_0}}{2 \pi (T-x) \sqrt{1-x}} ~, \label{eq:siggapped}
\eea
where the last equality follows from partial fractions and identities \eref{intprinc3} and \eref{intprinc4}.  

\paragraph{Conditions for gap formation}
In the gapped phase,~\eqref{eq:siggapped} is positive when $\CR > 2$. The condition $-1 \le a_0 \le 1$ for the expression~\eqref{asol} further leads to 
\bea
1<T \leq T_c \qquad \text{or} \qquad t_c \leq t < 1~.
\eea
Notice that these conditions are complementary to those of the gapless phase above.\footnote{In the critical case $\CR>2$ and $t=t_c$, we found that both solutions~\eqref{sigmagpl} and~\eqref{eq:siggapped} are valid, and in this case the two expressions indeed match.} We have thus identified the LO solution for all acceptable parameter values.

\subsubsection{The free energy}

In order to compute the LO terms for the free energy in~\eqref{freeenA} we need to compute $A$ of~\eqref{constantA} and $I$ of~\eqref{freeenc} at LO in the Veneziano limit.

\subsubsection*{Gapless phase} 
The LO result for $A$ can be calculated from its definition~\eqref{eq:poteq2}. The calculation can be simplified by using the identity~\eqref{Aexact},
which, by its derivation, holds for \emph{any} potential $\tv(x)$ and charge density $\sigma$ having the support $[a,b]$.
Applying the identity to the LO equation~\eqref{eq:poteq2} with $a=-1$ and $b=1$, we obtain
\bea \label{LOAgpl}
 A_0 &= \frac{\CR}{\pi} \int_{-1}^1 \frac{\log (T-x)}{\sqrt{1-x^2}}\,\ud x + 2 \log 2 \nonumber\\
   &= -\CR \log (2T-2\sqrt{T^2-1}) + 2 \log 2= -\CR \log (2t) + 2 \log 2~.
\eea
The LO integral $I_0$ is computed directly from \eref{freeenc}, giving
\bea \label{LOIgpl}
I_0 &= \int_{-1}^1 \tv_0(x) \sigma_0(x)\, \ud x \nn \\
  &= \frac{(\CR-2)\CR}{2}\log\left(2T-2\sqrt{T^2-1}\right) - \frac{\CR^2(T-t)}{\sqrt{T^2-1}} \log\left(\frac{1}{2\sqrt{T-1}}+\frac{1}{2\sqrt{T+1}}\right)\nn \\
  &= - \CR \log\left(2 t\right)+\CR^2 \log\left(1-t^2\right)~,
\eea
where we used the integral identities~\eqref{integralA2} and~\eqref{integralA3t}.

\subsubsection*{Gapped phase} 
Setting $a=a_0$ and $b=1$ in \eref{Aexact}, we obtain
\bea \label{LOAgpp}
 A_0 &= \frac{\CR}{\pi} \int_{a_0}^1 \frac{\log(T-x)}{\sqrt{(1-x)(x-a_0)}}\,\ud x - 4 \log \left(\frac{\sqrt{1-a_0}}{2}\right) \nonumber\\
 &= 2(\CR-1) \log (\CR-1)-2(\CR-2) \log(\CR-2)  +(\CR-2)\log  (T-1) \nonumber\\
&= \CR \log(\CR-1)+(\CR-2)\log\left[\frac{(\CR-1)(1-t)^2}{2(\CR-2)^2t}\right]~.
\eea
The integral $I_0$ is computed as in the gapless phase by using the definition \eref{freeenc}:
\bea \label{LOIgpp}
 I_0 &= \int_{a_0}^1 \tv_0(x) \sigma_0(x) \, \ud x\nn \\\nn 
 &=  -\frac{\CR(\CR-2)\sqrt{T-a_0}}{\sqrt{T-1}} \log\left(\frac{1}{2\sqrt{T-a_0}}+\frac{1}{2\sqrt{T+1}}\right) - \CR(\CR-2)\, \log\frac{\sqrt{T-a_0}+\sqrt{T-1}}{2} \\
 &= \CR(2\CR-1)\log 2+\CR \log\frac{(1-t)^2}{4 t}+\CR(\CR-2)\log\frac{\CR-2}{\CR}-2 \CR(\CR-1)\log\frac{2\CR-2}{\CR}~.
\eea

\subsection{Linear perturbation to the free energy} \label{app:NLO}

Let us then discuss how the NLO correction to the free energy can be obtained. Expanding~\eqref{freeen2} up to NLO we find that (see also, e.g., \cite{ChenLawrence})
\bea
 \CH = \frac{1}{2}\left(A_0+I_0\right) + \frac{1}{n}\Delta_\CH^{(-1)} + \morder{\frac{1}{n^2}}~,
\eea
where
\bea \label{NLOdef}
 \Delta_\CH^{(-1)}  &=- \int_a^1   \sigma_0(x) \Delta_\tv^{(-1)}(x)\, \ud x\nonumber\\
           &= -\alpha \int_a^1 \sigma_0(x) \log (1-x)\, \ud x - \beta \int_a^1  \sigma_0(x) \log (1+x)\, \ud x \ .
\eea
Here $a$ equals $-1$ in the gapless phase and $a_0$, as given by \eref{asol}, in the gapped phase. Notice that this expression is independent of the NLO solution $\Delta_\sig^{(-1)}(x)$. This was expected: $\sig_0(x)$ extremizes the Hamiltonian at LO, and therefore \emph{any} variation around it leaves the Hamiltonian unchanged at linear order. Therefore, only the LO solution to the equilibrium condition is required to compute the variation of the Hamiltonian $\Delta_\CH^{(-1)}$, as is usual in perturbation theory.

\paragraph{The gapless phase} The NLO corrections to the free energy are obtained by direct integration of~\eqref{NLOdef}. We need to evaluate the following integrals with $\sigma_0(x)$ given by the gapless phase case of \eref{sigmagpl}:
\bea
J_{\pm}:= -\int_{-1}^1  \sigma_0(x) \log (1\pm x)\,\ud x &= \frac{1}{2\pi} \int_{-1}^1\left(2-\CR + \frac{\CR \sqrt{T^2-1}}{T-x}\right)\frac{ \log(1\pm x)}{\sqrt{1-x^2}}\,\ud x \nonumber\\
&= -\frac{\CR-2}{2}\log 2 -\frac{\CR}{2}\,\log\left(\frac{T\pm 1 }{T+\sqrt{T^2-1}}\right) \\\nonumber
&=  \log 2 - \CR \log\left(1 \pm t \right)~.
\eea
Therefore, from \eref{NLOdef}, the NLO correction to the free energy is given by
\bea \label{NLOHgpl}
\Delta_\CH^{(-1)} 
&= \alpha J_- + \beta J_+\nn \\ 
&= {\alpha}\left[ \log 2 - \CR \log\left(1 - t \right)\right]+\beta \left[ \log 2 - \CR \log\left(1 + t \right)\right]~.
\eea

\paragraph{The gapped phase} In the gapped phase we proceed by direct integration as well. The relevant integrals are as follows with $\sigma(x)$ given by the gapped phase case of \eref{eq:siggapped}:
\bea
 \CI_\pm &:= - \int_{a_0}^1  \sigma(x) \log (1\pm x)\,\ud x  \nn \\
 &= -\frac{\CR-2}{2\pi} \int_{a_0}^1\frac{1}{(T-x)} \sqrt{\frac{x-a_0}{1-x}} \, \log(1\pm x)\, \ud x \nn \\
 & = -\frac{\CR-2}{2\pi} \int_{a_0}^1\left(\frac{T-a_0}{T-x}-1\right)\frac{\log(1\pm x)}{\sqrt{(1-x)(x-a_0)}}\, \ud x~.
\eea
The relevant integral identities are \eref{B2pm} and \eref{B3pm}.   With the aid of these identities, we obtain
\bea \label{Jpmdef}
 \CI_- &= \frac 12 A_0 - \frac{\CR}{2}  \log(T-1) \nonumber\\
     &=-\CR\, \log\frac{\CR-2}{\CR-1}-\log\frac{(\CR-1) (1-t)^2}{2 (\CR-2)^2 t}~, \\\nonumber
 \CI_+ &= \frac 12 A_0 -\frac{\CR}{2} \log(T+1)+ \frac{\CR}{2}\log\frac{\sqrt{\frac{2}{1+a_0}}-\frac{\CR-2}{\CR}}{\sqrt{\frac{2}{1+a_0}}+\frac{\CR-2}{\CR}}- \frac{\CR-2}{2}\log\frac{\sqrt{\frac{2}{1+a_0}}-1}{\sqrt{\frac{2}{1+a_0}}+1}~,
\eea
where $A_0$ is given in~\eqref{LOAgpp}.  Substituting $a_0$ from \eref{asol} and simplifying, we obtain the correction to the free energy from \eref{NLOdef} as
\bea \label{NLOHgpp}
\Delta_\CH^{(-1)} &= \alpha \CI_- + \beta \CI_+ \nn \\
&= - \frac{1}{2} \Bigg[ (\CR-2)\beta \log2 +2 (\alpha +\beta )(\CR-2) \log(\CR-2) -2\CR (\alpha +\beta ) \log(\CR-1) \nn \\
& \qquad \quad -2(\CR-2)\beta \log (\CR+S-2) +2 \alpha \log\{(\CR-1) (T-1)\} \nn \\
& \qquad \quad + \CR \beta  \log\{ 1+\CR (\CR+S-T-1)+T \} \Bigg]~,
\eea
where $S=\sqrt{\CR^2-2(\CR-1) (1+T)}$.

\subsection{Discussion}
The free energy in each phase can now be obtained by using
\bea \label{Ffinal}
 F = \frac{n^2}{2}\left(A_0+I_0\right) + n \Delta_\CH^{(-1)}+\morder{n^0}
\eea
and the results in~\eqref{LOAgpl},~\eqref{LOIgpl}, and~\eqref{NLOHgpl} [in~\eqref{LOAgpp},~\eqref{LOIgpp}, and~\eqref{NLOHgpp}] for the gapless [gapped] phase. 

As a consistency check, all results obtained in this Section match with those of Section~\ref{sec:pertsolutions}. In particular,
\bi
 \item The charge densities~\eqref{sigmagpl} and~\eqref{eq:siggapped} equal the leading order terms of~\eqref{densitygapless} and~\eqref{densitygapped}, respectively.
 \item The results for the free energy, as obtained from~\eqref{Ffinal}, match with~\eqref{Fgapless} and~\eqref{finalfreeen} in the gapless and gapped phases, respectively.
\ei
Thus, we also see that inserting the free energy of~\eqref{Ffinal} in the relation~\eqref{FandD} reproduces the key results of this article, given above in~\eqref{HSgapless} and in~\eqref{HSCngap}--\eqref{HSDngap}.

\section{Phase transition}
Having been discussing the two phases of the moduli space in the preceding sections, we now examine the transition between these two phases. Let us assume throughout this section that $\CR >2$ so that the transition is present.  We emphasise that, according to \eref{confwindow}, any value of $\CR$ within the conformal window gives rise to the phase transition.

Notice that the large $n$ limit is also required for the transition to take place: for any finite $n$ the integral $\CD_G$ in~\eqref{HS} is an analytic function of its parameters, and hence no phase transitions are expected. At finite but large $n$, we thus expect that the system is well described by the (leading order) results in the Veneziano limit except for very close to the transition point, where the phase transition is smoothed out. We can use the next-to-leading corrections, which we have calculated above, to estimate the scale where the smooth-out takes place. Let us now discuss in detail how this works.

\subsection{The gap width near the transition point}   
We first analyse the width of the gap in the gapped phase near the transition point.  Indeed, this width can be regarded as the {\it order parameter} of our system: if it vanishes (at leading order in the Veneziano limit), the system is in the gapless phase, otherwise the system is in the gapped phase. 

Consider the system in the gapped phase.  The gap spans the interval $[-1,a]$ and hence the gap width is $L:= a - (-1)=a+1$.  
Using \eref{ansatzeab}, \eref{a0b2}, and \eref{eq:a1}, we obtain
\bea
L \sim 1+ \frac{\CR^2-4 (\CR-1) T}{(\CR-2)^2} -\frac{2 \CR^2 (T-1) [S (\alpha +\beta )-(\CR-2) \beta ]}{n(\CR-2)^3 S} + \morder{n^{-2}}~.
\eea
Define
\bea
\epsilon :=T_c - T~, \qquad T< T_c~\text{\ in the gapped phase}
\eea
and consider the limit $T$ approaches $T_c$ from below or $\epsilon \rightarrow 0^+$.  We obtain
\bea
L &\sim \frac{4(\CR-1)}{(\CR-2)^2} \epsilon + \frac{1}{n} \Bigg[ \frac{\CR^2 \beta }{\sqrt{2} (\CR-1)^{3/2}} \frac{1}{\sqrt{\epsilon}} -\frac{\CR^2 (\alpha +\beta )}{(\CR-2) (\CR-1)} \nn \\
& \quad -\frac{\sqrt{2} \CR^2 \beta }{(\CR-2)^2 \sqrt{\CR-1}}  \sqrt{\epsilon } + \morder{\epsilon} \Bigg] + \morder{n^{-2}}~, \qquad \epsilon \rightarrow 0^{+}~. \label{gapsize}
\eea
We encounter here a double scaling problem, since $n$ and $\epsilon$ can be varied independently.
We can identify two regimes:
\ben
\item {\bf The case that $n^{-1} \epsilon^{-1/2} \ll \epsilon$, or equivalently $n \epsilon^{3/2} \rightarrow \infty$.}  The first term of \eref{gapsize} is dominant.  The order parameter near the transition point is therefore 
\bea
L_0 = \frac{4(\CR-1)}{(\CR-2)^2} \epsilon~. \label{orderparam1}
\eea  
 \item {\bf The case that $\epsilon \ll n^{-1} \epsilon^{-1/2}$, or equivalently $n \epsilon^{3/2} \rightarrow 0$.} The second term of \eref{gapsize} is dominant over the first term, suggesting that the perturbation series (in $1/n$) diverges. We interpret this as a signal of the phase transition having disappeared, so that the gap size $L$ behaves smoothly over the transition region.\footnote{We have verified this numerically for the Coulomb gas system of the $Sp(n)$ gauge group defined in Section~\ref{sec:exactv}.}
\een
The scale at which the phase transition is smoothed out is thus $\epsilon \sim n^{-2/3}$.

\subsection{Free energy at the phase transition}  
Let us consider the difference between the free energies of the two phases:
\bea
\delta F  := F_{\text{gapped}}-F_{\text{gapless}}
\eea
in the limit
\bea
\epsilon:= T_c-T \rightarrow 0^+ \quad  \text{as}\quad  n \rightarrow \infty~.  
\eea
Using \eref{Fgapless} and \eref{finalfreeen}, we obtain
\bea \label{explicitDF}
\delta F &\sim \left[ \frac{8(\CR-1)^5}{3 \CR^4(\CR-2)^4} \epsilon^3 -\frac{12(\CR-1)^6 [\CR(\CR-2)+2]}{\CR^6 (\CR-2)^6} \epsilon ^4 +\morder{\epsilon^5} \right] n^2 \nn \\
& \qquad + \Bigg[ \frac{8 \sqrt{2} (\CR-1)^{5/2} \beta}{3 \CR^2 (\CR-2)^2} \epsilon^{3/2}  + \frac{2(\CR-1)^3 (\alpha+\beta) }{\CR^2 (\CR-2)^3} \epsilon^2 + \morder{\epsilon^{5/2}}   \Bigg] n + \morder{n^0}~.
\eea
Let us now consider $\delta F$ in the cases discussed in the preceding subsection:
\ben
\item {\bf The case $n \epsilon^{3/2} \rightarrow \infty$}  The leading contribution to \eref{explicitDF} is given by
\bea
\frac{\delta F}{n^2} \sim \frac{8(\CR-1)^5}{3 \CR^4(\CR-2)^4} \epsilon^3 =\frac{(\CR-2)^2 (\CR-1)^2 }{24 \CR^4} L_0^3~. \label{3rdorder}
\eea
where the order parameter $L_0$ is given in \eref{orderparam1}. This case exhibits the {\it third order phase transition}, in a similar fashion to \cite{Gross:1980he, Wadia:2012fr}.
\item {\bf The case $n \epsilon^{3/2} \rightarrow 0$} 
The terms on the second row of \eref{explicitDF} are leading with respect to the ones on the first row, which again suggest that the perturbative results cannot be trusted. We expect that there is no phase transition but a smooth cross-over in this limit.
\een

\acknowledgments
Y.~C., N.~J., and M.~J. thank the Max Planck Institute for Physics for warm hospitalities while this work was in progress.  N.~M. wishes to express his sincere gratitude to Amihay Hanany for numerous discussions since 2008 that triggered his interest in this direction.  He also thanks the hospitality of Universidade de Santiago de Compostela, CERN Winter School 2013, ICTP Spring School 2013, and the Fizzi-Fazzi doublet (Fabio Apruzzi and Marco Fazzi) during his visit. The work of N.~M. is supported by a research grant of the Max Planck Society.  
N.~J. is supported by the MICINN and FEDER (grant FPA2011-22594) and the Spanish Consolider-Ingenio 2010 Programme CPAN (CSD2007-00042). N.J. is also supported by the Juan de la Cierva program. The work of M.J. was in part supported by grants
 PERG07-GA-2010-268246, PIEF-GA-2011-300984, the EU program ``Thales'' ESF/NSRF 2007-2013, and by the European Science
Foundation ``Holograv" (Holographic methods for strongly coupled systems) network.
It has also been co-financed by the European Union (European Social Fund, ESF) and
Greek national funds through the
 Operational Program ``Education and Lifelong Learning'' of the National Strategic
 Reference Framework (NSRF) under
 ``Funding of proposals that have received a positive evaluation in the 3rd and 4th Call of ERC Grant Schemes''.

\appendix

\section{Integral identities} 
We collect the relevant integral identities used in preceding sections.  Many of them are derived and listed in Appendix of \cite{ChenMcKay}.

\subsection{Integrals for the charge density}
In the following, we assume that $x \in (a,b)$ and $T \notin (a,b)$.  The notation $\CP$ denotes the principal value. We have
\bea
\int_{a}^{b}\frac{\ud x}{(T \pm x)\sqrt{(b-x)(x-a)}}  &=\frac{\pi}{\sqrt{(T \pm a)(T \pm b)}}~. \label{intCndens} \\
\CP \int_{a}^{b} \frac{\ud y}{(x-y)\sqrt{(b-y)(y-a)}} &= 0~, \label{intprincipal} \\
\CP \int_a^b \frac{\sqrt{(b-y)(y-a)}}{(y-x)(y \pm T)} \ud y &= \pi \left( \frac{\sqrt{(T \pm a)(T \pm b)}}{T-x} -1 \right)~, \label{intprinc2} \\ 
\CP \int_a^b \frac{\sqrt{(b-y)(y-a)}}{y-x} \ud y &= \frac{1}{2} \pi (a+b-2x)~, \label{intprinc3} \\
\int_a^b \frac{\sqrt{(b-y)(y-a)}}{y \pm T} \ud y &= \frac{1}{2} \pi \left[ a+b \pm 2T \mp 2 \sqrt{(T \pm a)(T \pm b)} \right]~,  \label{intprinc4} \\
\CP \int_a^b \frac{1}{(y-x)} \sqrt{\frac{y-a}{1-y}} \ud y &= \pi~.\label{intprinc5}
\eea

\subsection{Integrals for the free energy} \label{app:intF}
Let us present the integral identities used in Section \ref{sec:freeenergy}.
\subsubsection*{The integrals $I^\pm_1$} 
We compute the following integrals:
\bea
I^\pm_1 &= \int_a^b \ud x \frac{\log (T-x) \sqrt{(b-x)(x-a)} }{(1 \pm x)} \nn \\
&= \int_a^b \ud x \frac{ \log (T-x)}{\sqrt{(b-x)(x-a)}} \frac{(b-x)(x-a)}{1\pm x}\nn \\
&= \int_a^b \ud x \frac{ \log (T-x)}{\sqrt{(b-x)(x-a)}} \left[ \mp (x-a)+(1\pm b)- \frac{(1\pm a)(1 \pm b)}{1 \pm x}\right]  \nn \\
&= \mp \int_a^b \ud x \sqrt{\frac{x-a}{b-x}} \log (T-x)~\ud x+ (1 \pm b)  \int_a^b \frac{ \log (T-x)}{\sqrt{(b-x)(x-a)}} ~\ud x \nn \\
& \qquad - (1 \pm a)(1 \pm b) \int_a^b  \frac{ \log (T-x)}{(1 \pm x)\sqrt{(b-x)(x-a)}}~ \ud x \nn \\
&= \mp A_1 + (1 \pm b) A_2 - (1 \pm a)(1 \pm b) A^{\pm}_3~,
\eea
where, using (6.1) and (6.5) of \cite{ChenMcKay}, we have
\bea
A_2 &=  \int_a^b \frac{ \log (T-x)}{\sqrt{(b-x)(x-a)}} ~\ud x  = 2 \pi \log \left( \frac{\sqrt{T-a} + \sqrt{T-b}}{2} \right)~, \label{integralA2} \\
A^\pm_3 &=  \int_a^b  \frac{ \log (T-x)}{(1 \pm x)\sqrt{(b-x)(x-a)}}~ \ud x  \nn \\
&=  \frac{\pi }{\sqrt{(1 \pm a)(1 \pm b)}}  \log \left(  \frac{ \pm (T \pm 1)^2 \mp (\sqrt{(T-a)(T-b)} \mp \sqrt{(1 \pm a)(1 \pm b)})^2}{(\sqrt{1\pm a}+\sqrt{1\pm b})^2}\right)~,
\eea
and using (6.1) and (6.4) of \cite{ChenMcKay}, we obtain
\bea \label{idenA1}
A_1=  \int_a^b \ud x \sqrt{\frac{x-a}{b-x}} \log (T-x)~\ud x 
 &= \int_a^b \frac{x-a}{\sqrt{(b-x)(x-a)}} \log (T-x) ~\ud x \nn \\
 &= \int_a^b \frac{x  \log (T-x) }{\sqrt{(b-x)(x-a)}}~\ud x - a \int_a^b  \frac{ \log (T-x)}{\sqrt{(b-x)(x-a)}}~ \ud x \nn \\
 &= \pi  \Bigg[ \frac{1}{2}(b+a) - T + \sqrt{(T-a)(T-b)}  \nn \\
 & \qquad + (b-a) \log \left( \frac{\sqrt{T-a}+ \sqrt{T-b}}{2} \right) \Bigg]~.
\eea

\subsubsection*{The integrals $I^{\pm \pm}_2$}
We compute the following integrals:
\bea
I^{- \pm}_2 &=  \int_a^b \ud x \log (1- x) \frac{\sqrt{(b-x)(x-a)}}{1 \pm x} \nn \\
&= (I^{\pm}_1)_{T=1}~, \\
I^{+ \pm}_2 &=  \int_a^b \ud x \log (1+ x) \frac{\sqrt{(b-x)(x-a)}}{1 \pm x} \nn \\
&=  (I^{\mp}_1)_{T=1,a \rightarrow -a, b \rightarrow -b}~.\eea

\subsubsection*{The integral $I_3$} 
We compute the following integral:
\bea
I_3 &=  \int_a^b \ud x \frac{\log (T-x) \sqrt{(b-x)(x-a)} }{(T-x)}  \nn \\
&= \int_a^b \ud x \frac{ \log (T-x)}{\sqrt{(b-x)(x-a)}} \frac{(b-x)(x-a) }{(T-x)}  \nn \\
&= \int_a^b \ud x \frac{ \log (T-x)}{\sqrt{(b-x)(x-a)}} \left[ (x-a)+(T-b) + \frac{(T-a)(T-b)}{x-T}\right]  \nn \\
&= \int_a^b \ud x \sqrt{\frac{x-a}{b-x}} \log (T-x)~\ud x+ (T-b)  \int_a^b \frac{ \log (T-x)}{\sqrt{(b-x)(x-a)}} ~\ud x \nn \\
& \qquad - (T-a)(T-b) \int_a^b  \frac{ \log (T-x)}{(T-x)\sqrt{(b-x)(x-a)}}~ \ud x \nn \\
&=: A_1 + (T-b) A_2 - (T-a)(T-b) \widetilde{A}_3~,
\eea
where
\bea \label{integralA3t}
\widetilde{A}_3 = \int_a^b  \frac{ \log (T-x)}{(T-x)\sqrt{(b-x)(x-a)}}~ \ud x &= - \frac{2 \pi}{\sqrt{(T-a)(T-b)}} \log \left(\frac{1}{2 \sqrt{T-a}} + \frac{1}{2 \sqrt{T-b}}  \right)~.
\eea

\subsubsection*{The integrals $I^{\pm}_4$}
We compute the following integrals:
\bea
I^{\pm}_4 &=  \int_a^b \ud x \log (1 \pm x) \frac{\sqrt{(b-x)(x-a)}}{T-x} \nn \\
&= \int_a^b \ud x \frac{\log (1\pm x)}{\sqrt{(b-x)(x-a)}}   \left[ (x-a)+(T-b) + \frac{(T-a)(T-b)}{x-T}\right]  \nn \\
&= \int_a^b  \sqrt{\frac{x-a}{b-x}} \log (1\pm x) \ud x + (T-b)  \int_a^b \frac{ \log (1\pm x)}{\sqrt{(b-x)(x-a)}} ~\ud x \nn \\
& \qquad - (T-a)(T-b) \int_a^b  \frac{\log (1\pm x)}{(T-x)\sqrt{(b-x)(x-a)}}~ \ud x \nn \\
&= B^{(1)}_{\pm}  + (T-b) B^{(2)}_{\pm} - (T-a)(T-b) B^{(3)}_{\pm}~,
\eea
where, from identities (6.6) and (6.9) of \cite{ChenMcKay}, we have
\bea
B^{(2)}_{\pm} & =\int_a^b \ud x \frac{\log (1 \pm x)}{\sqrt{(b-x)(x-a)}} =  2 \pi \log \left( \frac{\sqrt{1 \pm a} + \sqrt{1 \pm b}}{2} \right)~,  \label{B2pm} \\
B^{(3)}_{\pm} & =\int_a^b \ud x \frac{\log (1 \pm x)}{(T-x)\sqrt{(b-x)(x-a)}}  \nn \\
&= \frac{\pi }{\sqrt{(T-a)(T-b)}}  \log \left(  \frac{ \pm (T \pm 1)^2 \mp (\sqrt{(T-a)(T-b)} \mp \sqrt{(1 \pm a)(1 \pm b)})^2}{(\sqrt{T - a}+\sqrt{T - b})^2}\right)~, \label{B3pm}
\eea
and from \eref{idenA1}, we obtain
\bea
B^{(1)}_{\pm} &= \int_a^b \ud x  \sqrt{\frac{x-a}{b-x}} \log (1 \pm x)  \nn \\
& = \pi  \Bigg[ \frac{1}{2}(b+a) \pm 1 \mp \sqrt{(1 \pm a)(1 \pm b)}  + (b-a) \log \left( \frac{\sqrt{1\pm a}+ \sqrt{1 \pm b}}{2} \right) \Bigg]~.
\eea

\section{Method of conformal mappings and mirror charges} \label{app:confmap}
In this section, we present an alternative method to find the equilibirium charge density $\sigma(x)$ of the Coulomb gas, which makes use of conformal mappings and mirror charges. This method could be used to derive the results discussed in the Sections~\ref{sec:exactv}--\ref{sec:pertanalysis}. The strength of this method is that the free energy, which is required to derive the saddle-point approximation for the Hilbert series, can be calculated without solving the equilirium condition explicitly. As an example, we apply it to the calculation of Section~\ref{sec:pertanalysis}, and solve the free energy at leading order.

\subsection{Solving the complex potential}
We construct the complex potential
\bea \label{eq:Vydef}
\Phi(x) = \frac{1}{2} \CR \log(x-T)-  \int_{a}^1  \ud y ~ \sigma_0(y)  \log(x-y)  \ ,
\eea
where the LO charge density $\sigma_0(x)$ is fixed by requiring that the real part $\re~\Phi(x)$ is constant
\bea \label{AApp}
\re\left[\Phi(x) \right]= \frac{1}{2} A_0
\eea
on the support of the charge density $[a,1]$.  
By comparing~\eqref{eq:Vysol} to the definition~\eqref{eq:poteq2} we see that~\eqref{AApp} indeed matches with the LO constant $A_0$ of Section~\ref{sec:pertanalysis}.
The lower limit $a$ (with $-1<a<1$) is to be determined subsequently.

We start by applying the conformal map
\be \label{eq:fdef}
 x \quad \mapsto \quad  z = f(x) := \sqrt{\frac{x-1}{x-a}}
\ee
which maps the exterior of the ``conductor'' $[a,1]$ in the compactified complex plane to the right half-plane. The point charge $-\CR/2$ at $x=T$ is mapped to $z=Z$, where 
\bea 
Z:=f(T) = \sqrt{\frac{T-1}{T-a}} \qquad \text{and} \qquad 0 \leq Z \leq 1~.  \label{defZ}
\eea 
Notice also that there is a point particle with the charge $\CR/2-1$ at $x=\infty$, which is mapped to $z=1$. The solution for which the real part of the potential is constant on the image of the conductor, the imaginary axis, is now obtained by placing mirror charges in the left half-plane and reads\footnote{The complex potential of a point charge $e$ located at the point $z_0$ in two dimensions is given by $\widehat{\Phi}(z)= -e \log (z-z_0)$.} 
\bea \label{eq:Vzsol}
\widehat{\Phi}(z) = \frac{1}{2} \CR \log \left( \frac{z-Z}{z+Z} \right) -\left(\frac{1}{2}\CR-1\right) \log \left( \frac{z-1}{z+1} \right) + \Phi_0~. 
\eea
This solution is the unique one having the desired analytic structure, up to irrelevant branch choices and the value of the constant $\Phi_0$. 
Mapping back to the $x$-plane, we find the solution
\bea \label{eq:Vysol}
\Phi(x) &= \widehat{\Phi}\left( f(x) \right) \nn \\
&=  \frac{1}{2}\CR \log \frac{\sqrt{\frac{x-1}{x-a}}-\sqrt{\frac{T-1}{T-a}}}{\sqrt{\frac{x-1}{x-a}}+\sqrt{\frac{T-1}{T-a}}} -\left(\frac{1}{2}\CR-1\right) \log \frac{\sqrt{\frac{x-1}{x-a}}-1}{\sqrt{\frac{x-1}{x-a}}+1} + \Phi_0~.
\eea

\subsection{The free energy}
Evaluating the real part of Eq.~\eqref{eq:Vysol} on the conductor, say at $x=1$, we find that the constant in~\eqref{AApp} is
\bea
A_0 = 2\, \re \left[ \Phi( x=1) \right] = 2\, \re \Phi_0~.
\eea
The value of $\re \Phi_0$, and thus also the constant $A$, is found by comparing the large $x$ asymptotics of the solution~\eqref{eq:Vysol} to the definition~\eqref{eq:Vydef}. In this limit, Eq. \eref{eq:Vydef} gives
\bea
\re \Phi(x) = \left(\frac{1}{2}\CR-1\right) \log(x) + \morder{\frac{1}{x}}~, \qquad x \rightarrow \infty~.
\eea  
whereas Eq.~\eqref{eq:Vysol} gives
\bea
\re \Phi(x) = \frac{1}{2} \CR \log \frac{1-\sqrt{\frac{T-1}{T-a}}}{1+\sqrt{\frac{T-1}{T-a}}} -\left(\frac{1}{2}\CR-1\right) \log \left( \frac{1-a}{4x} \right) +  \frac{1}{2} A_0 + \morder{\frac{1}{x}}~, \qquad x \rightarrow \infty~.
\eea
Equating these two expressions, we have
\bea
 A_0 = (\CR-2) \log \left( \frac{1-a}{4} \right) - 
 \CR \log \frac{1-\sqrt{\frac{T-1}{T-a}}}{1+\sqrt{\frac{T-1}{T-a}}} ~. \label{U0sol}
\eea

On the other hand, at LO the integral $I$ of~\eqref{freeenc} is given by
\bea
 I_0 = \CR \int_a^1 \log(T-x)\, \sigma_0(x)\,\ud x = -\CR \lim_{x \to T} \re \left[\Phi(x) - \frac{1}{2}\CR  \log(x-T)\right]~,
\eea
as seen from the definition~\eqref{eq:Vydef}. Inserting here the solution~\eqref{eq:Vysol} gives
\bea \label{I0solapp}
 I_0 = -\frac{\CR}{2} \left\{ (2\CR-2)\log\frac{1-a}{4} - \CR \log\left[(T-1)(T-a)\right]+(2-2\CR) \log\frac{1-\sqrt{\frac{T-1}{T-a}}}{1+\sqrt{\frac{T-1}{T-a}}}\right\}~.
\eea
The result for the LO free energy is thus proportional to the sum of~\eqref{U0sol} and~\eqref{I0solapp}.

\subsection{The solution for the gapless phase}
The solution for the gapless phase is now given by setting $a=-1$ in~\eqref{U0sol} and~\eqref{I0solapp}:
\bea
A_0 
 &= 2 \log 2 - \CR \log(2t)~, \label{app:Agapless} \\
I_0 &= - \CR \log(2t) + \CR^2 \log \left(1-t^2\right)~, \label{app:Igapless}
\eea
where we used \eref{Tandt} to replace $T$ by $t$. Observe that the results coincide with~\eqref{LOAgpl} and~\eqref{LOIgpl}.

\subsection{The solution for the gapped phase}
In order to determine the solution for the gapped phase, we need to determine the value of $a$.  To this end, it is necessary to study the behaviour of the charge density near its lower edge. 

Since conformal mappings conserve charge, it is useful to study the density in the $z$-plane, where $x=a$ is mapped to $z = \infty$ and $x=1$ is mapped to $z=0$. We interpret the left half-plane as the conductor with a constant potential, whereas in the right half-plane the potential is given by Eq.~\eqref{eq:Vzsol}. The charge density is the discontinuity of the electric field at the surface,
\bea
 \widehat{\sigma}(\zeta)  = -\frac{1}{2\pi } \left[ \frac{\ud \widehat{\Phi}(z)}{\ud z} \right]_{z=i \zeta} = \frac{Z [\CR - ( \CR-2) Z] +\zeta^2 [\CR Z- (\CR-2)]}{2\pi(1 + \zeta^2) (\zeta^2 + Z^2)}~, \label{sigmazeta}
\eea
where $\zeta = \im z$ and $Z$ is given by \eref{defZ}.

\paragraph{Conditions for gap formation} 
Let us assume that we are in the gapless phase ($a=-1$). The gap is formed when the charge density vanishes at $x=-1$, which is mapped to $\zeta = \pm \infty$.
As $\zeta \to \pm \infty$, the charge density behaves as
\bea
\widehat{\sigma}(\zeta) \sim \zeta^{-2}[\CR Z- (\CR-2)] \ , \qquad \zeta \rightarrow \infty \label{asympdensity}
\eea
Then, requiring \eref{asympdensity} to be non-negative yields
\bea
Z = \sqrt{\frac{T-1}{T+1}} = \frac{1-t}{1+t}  \ge \frac{\CR-2}{\CR}~,
\eea
where we used the definitions of $Z$ and $T$ from \eref{defZ} and \eref{Tandt}. This equality results in the same conditions for the gapless phase as in Sections~\ref{sec:pertsolutions} and~\ref{sec:pertanalysis}, $\CR \le 2$ or $ 0 \le t \le t_c = 1/(\CR-1)$. In the opposite case, the system is in the gapped phase.

\paragraph{The lower endpoint of the conductor in the gapped phase}
To solve the location of the lower endpoint $a=a_0$ of the charge density in the gapped phase (with $-1<a_0<1$), we require that the leading term of the charge density \eref{asympdensity} vanishes at the endpoint. This gives
\bea
Z= f(T) = \sqrt{\frac{T-1}{T-a_0}} = \frac{\CR-2}{\CR}~, 
\eea
from which the value of $a_0$ can be solved:
\bea \label{eq:asol}
 a_0= \frac{\CR^2-4 (\CR-1) T}{(\CR-2)^2}~. 
\eea

\paragraph{The free energy in the gapped phase}
The free energy in the gapped phase can now be obtained by inserting the solution~\eqref{eq:asol} to equations~\eqref{U0sol} and~\eqref{I0solapp}. It is easy to check that the result agrees with~\eqref{LOAgpp} and~\eqref{LOIgpp}.

\subsection{Charge densities} 
As a byproduct of the method, we can also solve the charge density on the $x$-plane by applying the conformal map \eqref{eq:fdef} to \eref{sigmazeta}. Let $x$ be a coordinate on support $[a,1]$ of the density $\sigma(x)$.  We obtain\footnote{Since each $x \in (-1,1)$ corresponds to two values on the imaginary $z$-axis, there is an extra factor of two with respect to the formula~\eqref{sigmazeta}. The limiting procedure is necessary in order to avoid the branch cut of $f(x)$.}
\bea \label{eq:rhoy1}
\sigma_0(x) &:= -\frac{i}{\pi} \lim_{\epsilon\to 0^+}\Phi'(x+i\epsilon) 
=  2 i \lim_{\epsilon\to 0^+} f'(x+i\epsilon)\, \widehat{\sigma}(-if(x+i\epsilon))~.
\eea
The explicit results for the charge densities can readily be computed:
\bea \label{eq:rhogpp}
\sigma_0(x) = \begin{cases} 
\frac{(2-\CR) (T-x)+\CR \sqrt{T^2-1}}{2 \pi  (T-x) \sqrt{1-x^2}}~,    & x \in [-1,1] \qquad \text{(gapless phase)}~, \\ 
\frac{(\CR-2)\sqrt{x-a_0}}{2\pi(T-x)\sqrt{1-x}}~,  & x \in [a_0,1] \hspace{1cm} \text{(gapped phase)}~.  
\end{cases}
\eea
These results are in accordance with~\eqref{sigmagpl} and~\eqref{eq:siggapped}.

\bibliographystyle{ytphys}
\bibliography{ref}

\end{document}